\documentclass[12pt]{iopart}

\usepackage{graphicx}
\usepackage{amssymb}
\usepackage{color}
\usepackage{epstopdf}

\begin{document}

\newcommand{\beq}{\begin{equation}}
\newcommand{\eeq}{\end{equation}}
\newcommand{\barr}{\begin{eqnarray}}
\newcommand{\earr}{\end{eqnarray}}
\newcommand{\andy}[1]{ }
\newcommand{\bmsub}[1]{\mbox{\boldmath\scriptsize $#1$}}
\def\bra#1{\langle #1 |}
\def\ket#1{| #1 \rangle}
\def\sinc{\mathop{\text{sinc}}\nolimits}
\def\cV{\mathcal{V}}
\def\cH{\mathcal{H}}
\def\cT{\mathcal{T}}
\def\cP{\mathcal{P}}
\def\e{\mathrm{e}}
\def\i{\mathrm{i}}
\def\d{\mathrm{d}}
\renewcommand{\Re}{\mathop{\mathrm{Re}}\nolimits}
\renewcommand{\Im}{\mathop{\mathrm{Im}}\nolimits}

\newcommand{\REV}[1]{\textbf{\color{red}#1}}
\newcommand{\BLUE}[1]{\textbf{\color{blue}#1}}
\newcommand{\GREEN}[1]{\textbf{\color{green}#1}}

\title[Classical Statistical Mechanics Approach to Multipartite Entanglement ]{Classical Statistical Mechanics Approach to Multipartite Entanglement }

\author{P. Facchi $^{1,2}$, G. Florio$^{3,2}$, U. Marzolino$^{4}$, G. Parisi$^{5}$, S. Pascazio$^{3,2}$}

\address{$^1$Dipartimento di Matematica, Universit\`a di Bari, I-70125 Bari, Italy}
\address{$^2$Istituto Nazionale di Fisica Nucleare, Sezione di Bari, I-70126 Bari, Italy}
\address{$^3$Dipartimento di Fisica, Universit\`a di Bari, I-70126 Bari, Italy}
\address{$^4$Dipartimento di Fisica, Universit\`a di Trieste, and Istituto Nazionale di Fisica Nucleare, Sezione di Trieste, I-34014 Trieste, Italy}
\address{$^5$Dipartimento di Fisica, Universit\`{a} di Roma ``La Sapienza", Piazzale Aldo Moro 2, \\
Centre for Statistical Mechanics and Complexity (SMC), CNR-INFM, \\
and Istituto Nazionale di Fisica Nucleare, Sezione di Roma, 00185
Roma, Italy}

\begin{abstract}
We characterize the multipartite entanglement of a  system of $n$
qubits in terms of the distribution function of the bipartite
purity over balanced bipartitions. We search for maximally
multipartite entangled states, whose average purity is minimal,
and recast this optimization problem into a problem of statistical
mechanics, by introducing a cost function, a fictitious
temperature and a partition function. By investigating the
high-temperature expansion, we obtain the first three moments of the
distribution. We find that the problem exhibits frustration.
\end{abstract}


\maketitle

\section{Introduction}\label{sec:intro}

There is a profound diversity between quantum mechanical and
classical correlations. Schr\"{o}dinger \cite{Schr1,Schr2} coined
the term ``entanglement" to describe the peculiar connection that can exist between
quantum systems, that was first perceived by Einstein, Podolsky and
Rosen \cite{EPR} and has no analogue in classical physics.
Entanglement is a resource in quantum information science
\cite{entanglement,entanglementrev,h4} and is at the origin of many
unique quantum phenomena and applications, such as superdense coding
\cite{densecoding}, teleportation \cite{teleport} and quantum
cryptographic schemes
\cite{crypto0,crypto1,crypto2,crypto3}.

Much progress has been made in developing a quantitative theory of
entanglement \cite{entanglementrev,h4}. The \emph{bipartite}
entanglement between simple systems can be unambiguously defined in
terms of the von Neumann entropy or the entanglement of formation
\cite{wootters,entanglement,entanglement2}. On the other hand, an
exhaustive characterization of \emph{multipartite} entanglement is
more elusive \cite{entanglementrev,h4} and different definitions
\cite{multipart1,multipart2,multipart3,multipart4,Bergou}
often do not agree with each other, essentially because they tend to
capture different aspects of the phenomenon. More to this, a
complete evaluation of global entanglement measures bears serious
computational difficulties, because states endowed with large
entanglement typically involve exponentially many coefficients.

We proposed in \cite{statmech} that multipartite entanglement shares
many characteristic traits of complex systems and can therefore be
analyzed in terms of the probability density function of an
entanglement measure (say purity) of a subsystem over all (balanced)
bipartitions of the total system \cite{FFP}. A state has a large
\emph{multipartite} entanglement if its average bipartite
entanglement is large. In addition, if the entanglement distribution
has a small standard deviation, bipartite entanglement is
essentially independent of the bipartition and can be considered as
being fairly ``shared" among the elementary constituents (qubits) of
the system. Clearly, average and standard deviation are but the
first two moments of a distribution function. A full
characterization of the multipartite entanglement of a quantum state
must therefore take into account higher moments and/or the whole
distribution function, in particular if the latter is not bell
shaped or is endowed with unusual and/or irregular features.

The idea that complicated phenomena cannot be summarized in a single
(or a few) number(s), but rather require a large number of measures
(or even a whole function) is not novel in the context of complex
systems \cite{parisi} and even in the study of quantum entanglement
\cite{MMSZ}. In this article we shall pursue this idea even further
and shall study the bipartite and multipartite entanglement of a
system of qubits by making full use of the tools and techniques of
classical statistical mechanics: we shall explore the features of a partition
function, expressed in terms of the average purity of a subset of the qubits: this will be viewed as
a cost function, that plays the role
of the Hamiltonian. Interestingly, this approach brings to light the
presence of frustration in the system \cite{frustr}, highlighting the complexity
inherent in the phenomenon of multipartite entanglement.

This paper is organized as follows. We introduce notation and define
maximally bipartite and maximally multipartite entangled states in
Sec.\ \ref{sec:definitions}. Multipartite entanglement is
characterized in terms of the distribution function of bipartite
entanglement in Sec.\ \ref{sec:pdf}. The statistical mechanical
approach and the partition function are introduced in Sec.\
\ref{sec:partfun}. The high temperature expansion and its first three
cumulants are computed in Sec.\ \ref{sec:hight}. 
Section \ref{sec:concl} contains our conclusions and an outlook.

\section{From bipartite to multipartite entanglement}
\label{sec:definitions}

\subsection{Bipartite purity }

We consider an ensemble $S=\{1,2,\dots, n\}$ of $n$ qubits in the
Hilbert space $\mathcal{H}_S= (\mathbb{C}^2)^{\otimes n}$ and focus
on pure states
\begin{equation}
|\psi\rangle = \sum_{k\in {\mathbb{Z}}_2^n} z_k |k\rangle , \quad z_k \in
\mathbb{C}, \quad \sum_{k\in {\mathbb{Z}}_2^n} {|z_k|}^2 =1,
\label{eq:genrandomx}
\end{equation}
where $k=(k_i)_{i\in S}$, with $k_i\in {\mathbb{Z}}_2=\{0,1\}$, and
\begin{equation}
\label{eq:ki}
\ket{k}=\bigotimes_{i\in S} \ket{k_i}_i, \qquad
\ket{k_i}_i \in \mathbb{C}^2, \qquad \bra{k_i} k_j \rangle= \delta_{ij}  .
\end{equation}
For the sake of simplicity, in this paper we shall focus on pure
states of qubits and shall not discuss additional phenomena such as
bound entanglement \cite{boundent1,boundent2}. Consider a
bipartition $(A,\bar{A})$ of the system, where $A \subset S$ is a
subset of $n_A$ qubits and $\bar{A}=S\backslash A$ its complement,
with $n_A+n_{\bar{A}}=n$. We set $n_A
\leq n_{\bar{A}}$ with no loss of generality. The total Hilbert space
factorizes into
$\mathcal{H}_S=\mathcal{H}_A\otimes\mathcal{H}_{\bar{A}}$, with
$\mathcal{H}_A= \bigotimes_{i\in A} \mathbb{C}^2_i$, of dimensions
$N_A=2^{n_A}$ and $N_{\bar{A}}=2^{n_{\bar{A}}}$, respectively
($N_AN_{\bar{A}}=N$). As a measure of the \emph{bipartite}
entanglement between the two subsets, we consider the purity of
subsystem $A$
\begin{equation}
\pi_{A}=\tr_A\rho_{A}^2, \quad \rho_{A}=\tr_{\bar{A}}|\psi\rangle\langle\psi|,
\label{eq:puritydef}
\end{equation}
$\tr_{X}$ being the partial trace over $X=A$ or $\bar{A}$. We notice
that $\pi_{A}=\pi_{\bar{A}}$ and
\beq
\label{eq:purityconstraint}
1/N_A\le\pi_{A}\le 1.
\eeq
State (\ref{eq:genrandomx})  can be written according to the bipartition $(A,\bar{A})$ as
\begin{equation}
|\psi\rangle = \sum_{k\in {\mathbb{Z}}_2^n} z_k \ket{k_A}_A\otimes\ket{k_{\bar{A}}}_{\bar{A}} ,
\label{eq:genrandomxbip}
\end{equation}
where $k_A=(k_i)_{i\in A}$ and $\ket{l}_A=\bigotimes_{i\in A}\ket{l_i}_i\in\mathcal{H}_A$.
By plugging Eq.\ (\ref{eq:genrandomxbip}) into Eq.\
(\ref{eq:puritydef}) we obtain
\begin{equation}
\label{eq:rhoA}
\rho_A=  \sum_{k , l\in {\mathbb{Z}}_2^n}
z_{k} \bar{z}_{l}
\delta_{k_{\bar{A}},l_{\bar{A}}}
\ket{k_A} \bra{l_A}
\end{equation}
and
\begin{equation}
\pi_{A}= \sum_{k,k',l,l' \in {\mathbb{Z}}_2^n} z_{k} z_{k'}
\bar{z}_{l}\bar{z}_{l'} \delta_{k_A, l'_A}
\delta_{k'_A,l_A} \delta_{k_{\bar{A}}, l_{\bar{A}}} \delta_{k'_{\bar{A}},
l'_{\bar{A}}} ,
\label{eq:puritygeneral}
\end{equation}
which is a quartic function of the coefficients of the expansion
(\ref{eq:genrandomx}). If, for example, the system is partitioned
into two blocks of contiguous qubits  $(C,\bar{C})$,  namely
$C=\{1,2,\dots, n_A\}$, then
\beq
\pi_{C}= \sum_{l,l'\in {\mathbb{Z}}_2^{n_A}}\sum_{m,m'\in {\mathbb{Z}}_2^{n_{\bar{A}}}} z_{(l,m)}
\bar{z}_{(l', m)}z_{(l', m')}\bar{z}_{(l, m')},
\label{eq:purityref}
\eeq
where $(l,m)=(l_1,\dots,l_{n_A},m_1,\dots,m_{n_{\bar{A}}})\in {\mathbb{Z}}_2^n$.

\subsection{Minimal bipartite purity}
\label{sec:minimal}

For a given bipartition it is very easy to saturate the lower
bound $1/N_A$ of (\ref{eq:puritygeneral}). For example,
\begin{equation}
\label{eq:maxbi1}
z_k=N_A^{-1/2} \delta_{k_A,k_{\bar{A}}} ,
\end{equation}
which represents a \emph{maximally bipartite entangled  state}
\begin{equation}
\label{eq:maxbistate1}
|\psi\rangle = N_A^{-1/2} \sum_{l\in  {\mathbb{Z}}_2^{n_A}}\ket{l}_A\otimes\ket{l}_{\bar{A}} ,
\end{equation}
yields $\rho_A= \mathbf{1}/N_A$ and $\pi_A=1/N_A$.
In fact, the general minimizer is a maximally bipartite entangled
state whose Schmidt basis is not the computational basis, namely,
\begin{equation}
\label{eq:maxbi}
z_k=N_A^{-1/2} \sum_{l\in {\mathbb{Z}}_2^{n_A}} U^A_{k_A,l} U^{\bar{A}}_{k_{\bar{A}},l} ,
\end{equation}
where $U^A_{l,l'}=\bra{l_A} U^A  \ket{l'_A}$ with $U^A$ a local
unitary operator in $\mathcal{H}_A$ that transforms the
computational bases into the Schmidt one, that is
\begin{equation}
\label{eq:maxbistate}
|\psi\rangle = N_A^{-1/2} \sum_{l\in  {\mathbb{Z}}_2^{n_A}} U^A \ket{l}_A\otimes U^{\bar{A}} \ket{l}_{\bar{A}} .
\end{equation}
For this state, $\pi_{A}(\ket{\psi})=1/N_A$. The information
contained in a maximally bipartite entangled state with $n_A=n_{\bar{A}}$ is not locally accessible
by party $A$ or $\bar{A}$, because their partial density matrices
are maximally mixed, but rather is totally shared by them.

\subsection{Average purity and MMES}

Entanglement, in very few words, embodies the impossibility of
factorizing a state of the total quantum system in terms of the
states of its constituents. Most measures of bipartite entanglement
(for pure states) exploit the fact that when a (pure) quantum state
is entangled, its constituents do not have (pure) states of their own. This
is, for instance, what we did in the previous section. We wish to
generalize the above distinctive trait to the case of multipartite
entanglement, by requiring that this feature be valid for all
bipartitions.

Let $\ket{\psi}\in\mathcal{H}_S$ and consider the average purity
\cite{scott,MMES}
\begin{equation}
\label{eq:piave}
\pi_{\mathrm{ME}}^{(n)}(\ket{\psi})=\mathbb{E}[\pi_A] = \left(\begin{array}{l}n
\\n_A\end{array}\!\!\right)^{-1}\sum_{|A|=n_A}\pi_{A},
\end{equation}
where $\mathbb{E}$ denotes the expectation value, $|A|$ is the
cardinality of $A$ and the sum is over balanced bipartitions
$n_A=[n/2]$, where $[x]$ denotes the integer part of $x$. Since we are focusing on balanced
bipartitions, and any bipartition can be brought into any other
bipartition by applying a permutation of the qubits, the sum over balanced
bipartitions in (\ref{eq:piave}) is equivalent to a sum over the
permutations of the qubits. The quantity $\pi_{\mathrm{ME}}$
measures the average bipartite entanglement over all possible
balanced bipartitions and inherits the bounds
(\ref{eq:purityconstraint})
\begin{equation}
\label{eq:pmeconstraint}
1/N_A\le\pi_{\mathrm{ME}}^{(n)}(\ket{\psi})\le 1.
\end{equation}
The average purity introduced in Eq.\ (\ref{eq:piave}) is related to
the average linear entropy $S_L=\frac{N_A}{N_A-1}
(1-\pi_{\mathrm{ME}})$ \cite{scott} and extends ideas put forward in
\cite{multipart4,parthasarathy}.

A \emph{maximally multipartite entangled state} (MMES)
\cite{MMES} $\ket{\varphi}$ is a minimizer of
$\pi_{\mathrm{ME}}$,
\begin{eqnarray}
\label{eq:minimizer}
\pi_{\mathrm{ME}}^{(n)}(\ket{\varphi})= E_0^{(n)},  \\
\mathrm{with} \quad E_0^{(n)}=
\min \{\pi_{\mathrm{ME}}^{(n)}(\ket{\psi})\; | \; \ket{\psi}\in \mathcal{H}_S, \bra{\psi}\psi\rangle=1\}.
\nonumber\end{eqnarray}
The meaning of this definition is clear: the density
matrix of each subsystem $A\subset S$ of a MESS is as mixed as possible (given
the constraint that the total system is in a pure state), so that
the information contained in a MMES is as distributed as possible.

\subsection{Perfect MMES and the symptoms of frustration}
\label{sec:smalln}

For small values of $n$ one can tackle the minimization problem
(\ref{eq:minimizer}) both analytically and numerically. For
$n=2,3,5,6$ the average purity saturates its minimum in
(\ref{eq:pmeconstraint}): this means that purity is minimal
\emph{for all} balanced bipartitions. In this case we shall say that
the MMES is \emph{perfect}.

For $n=2$ (perfect) MMES are Bell states up to local unitary
transformations, while for $n=3$ they are equivalent to the GHZ
states \cite{GHZ}. For $n=4$ one numerically obtains $E_0^{(4)}=\min
\pi_{\mathrm{ME}}^{(4)}=1/3 > 1/4 =1 /N_A$
\cite{MMES,sudbery,sudbery2,higuchi}. For $n=5$ and 6 one can find
several examples of perfect MESS, some of which can be expressed in
terms of binary strings of coefficients [$z_k=\pm 1$ in Eq.\
(\ref{eq:genrandomx})].

The case $n=7$ is still open, our best estimate being $E_0^{(7)}
\simeq 0.13387 > 1/8 = 1 /N_A$. Most interestingly, perfect MMES  do
not exist for $n\geq 8$
\cite{scott}.
These findings are summarized in Table \ref{tab_mmes}. This brings
to light an intriguing feature of multipartite entanglement:
we observed in Sec.\
\ref{sec:minimal} that it is always possible to saturate the lower
bound in (\ref{eq:purityconstraint})
\begin{equation}
\label{eq:pmalb}
\pi_{A}^{(n)} = 1/N_A
\end{equation}
for a \emph{given} bipartition $(A,\bar A)$. However, in order to
saturate the lower bound
\begin{equation}
\label{eq:pmelb}
E_0^{(n)}=1/N_A
\end{equation}
in Eq.\ (\ref{eq:pmeconstraint}), it must happen that
(\ref{eq:pmalb}) be valid for any bipartition in the average
(\ref{eq:piave}). As we have seen, this requirement can be satisfied
only for very few ``special" values of $n$. For all other values of $n$ this is
impossible: different bipartitions ``compete" with each other, and
the minimum $E_0^{(n)}$ of $\pi_{\mathrm{ME}}^{(n)}$ is strictly
larger than $1/N_A$. We view this ``competition" among different bipartitions as a
phenomenon of frustration: it is already present for $n$ as small as
4  \cite{frustr}. (Interestingly, an analogous phenomenon exists also for ``Gaussian MMES", see \cite{lupo}.)

This frustration is the main reason for the difficulties one
encounters in minimizing $\pi_{\mathrm{ME}}$ in (\ref{eq:piave}).
Notice that the dimension of $\mathcal{H}_S$ is $N=2^n$ and the
number of partitions scales like $2^N$. We therefore need to define
a viable strategy for the characterization of MMES, when
$n\geq 8$.
\begin{table}[t]
\caption{Perfect MMES for different number $n$ of qubits. }
\label{tab_mmes}
\begin{center}
\begin{tabular}{|c|c|}
\hline
$n$ & perfect MMES \\
\hline
    2,3 & exist   \\
    4 & do not exist  \\
    5,6 & exist   \\
    7 & ?  \\
    $\geq 8$ & do not exist \\
    \hline
\end{tabular}
\end{center}
\end{table}

\section{Probability distribution of bipartite entanglement}
\label{sec:pdf}

We now introduce the distribution function of purity over all
bipartitions, $p(\pi_{A})$, that will induce a
probability-density-function characterization of multipartite
entanglement. For rather regular (i.e.\ bell-shaped) distributions
the first few moments already yield a good characterization: in
particular, the average will measure the amount of entanglement of
the state when the bipartitions are varied, while the variance will
quantify how uniformly is bipartite entanglement distributed among
balanced bipartitions.

The calculation of the properties of $\pi_A$ is particularly simple
for an important class of states. Consider the set
\begin{equation}
\label{eq:constraint}
C=\left\{(z_1,z_2,\ldots,z_N)\in\mathbb{C}^N | \sum_k 
|z_k|^2=1\right\} ,
\end{equation}
corresponding to normalized vectors in $\mathcal{H}_S$. This set is
left invariant under the natural action of the unitary group
$\mathcal{U}(\mathcal{H}_S)$. A \emph{typical state} is obtained by
sampling with respect to the action of
$\mathcal{U}(\mathcal{H}_S)$ on this set.
Typical states have been extensively
studied in the literature
\cite{lubkinrnd,lloydpagelsrnd,pagernd,zyczkowskirnd,scottcavesrnd,giraudrnd}
and can be (efficiently) generated by a chaotic dynamics
\cite{chaos,chaos1}.

For large $N$, the $\pi_A$'s have a bell-shaped
distribution over the bipartitions with mean and variance \cite{FFP}
\barr
\mu^{(n)} &=&\langle \pi_{A}\rangle_0
= \frac{N_A+N_{\bar{A}}}{N+1}, \label{eq:6a} \\
\sigma^2 &=&
\langle (\pi_{A}-\mu)^2\rangle_0=\frac{2(N_A^2-1)(N_{\bar{A}}^2-1)}{(N+1)^2(N+2)(N+3)} , \label{eq:6b}
\earr
respectively, where the brackets  $\langle \cdots \rangle_0$ denote
the average with respect to the unitarily invariant measure over
pure states
\begin{equation}
\d\mu_C(z)=\frac{(N-1)!}{\pi^N}\; \delta \! \left(1-\sum_k
|z_k|^2\right) \prod_k {\d z_k \d\bar{z}_k} ,
\label{eq:meastyp}
\end{equation}
induced by the Haar measure over $\mathcal{U}(\mathcal{H_S})$ through
the mapping $|\psi\rangle = \sum  z_j |j\rangle = U |\psi_0\rangle$,
for a given reference state $|\psi_0\rangle$ \cite{zyczkowskirnd}. Here $\d z_k \d\bar{z}_k= \d x_k \d y_k$, with $x_k= \Re z_k$ and  $y_k=\Im z_k$, denotes the Lesbegue measure on $\mathbb{C}$.

Given a state $\ket{\psi}\in\mathcal{H}_S$, the potential of multipartite
entanglement has the following expression in terms of
its Fourier coefficients $z_i$
\begin{eqnarray}
\pi_{\mathrm{ME}}
=  \sum_{k,k',l,l' \in {\mathbb{Z}}_2^n} \Delta(k, k'; l, l'
)\, z_{k}\, z_{k'}\,
\bar{z}_{l}\, \bar{z}_{l'}\, ,
\label{eq:pimeDelta}
\end{eqnarray}
with a \emph{coupling function}
\begin{equation}
\fl \quad \Delta(k,k';l,l'
) =\left(\!\!\begin{array}{c}n
\\n_A\end{array}\!\!\right)^{\!\!-1}\!\!\!\sum_{|A|=n_A}
\!\frac{1}{2} \left(
\delta_{k_{A}, l'_{A}}
\delta_{k'_{A},l_{A}}
 \delta_{k_{\bar{A}}, l_{\bar{A}}} \delta_{k'_{\bar{A}},
l'_{\bar{A}}}
+
\delta_{k'_{A}, l'_{A}}
\delta_{k_{A},l_{A}}
 \delta_{k'_{\bar{A}}, l_{\bar{A}}} \delta_{k_{\bar{A}},
l'_{\bar{A}}}
\right),
\label{eq:Deltadef}
\end{equation}
with $n_A=[n/2]$ (balanced bipartitions). The result follows by
plugging the expression (\ref{eq:puritygeneral}) of $\pi_A$  into
Eq.\ (\ref{eq:piave}), and by symmetrizing under the exchange
$k\leftrightarrow k'$ (or, equivalently, $A\leftrightarrow
\bar{A}$).   The coupling function $\Delta$ has the following
expression (see \ref{sec:appendixa} for details)
\begin{equation}\label{eq:deltag}
\Delta(k,k';l,l'
)= g\big((k\oplus l) \vee (k' \oplus l'), (k
\oplus l') \vee (k'\oplus l)
\big),
\end{equation}
where
\begin{equation}
 g(a,b
 ) = \delta_{a\wedge b,\,0} \; \hat{g}(|a|,|b|
 ),
\label{eq:gdef}
\end{equation}
with $|a|=\sum_{i\in S} a_i$, $|b|=\sum_{i\in S} b_i$, $a\oplus b=(a_i + b_i\; \mathrm{mod}\; 2)_{i \in S}$ is the XOR operation,  $a\vee b=(a_i + b_i - a_i b_i) _{i \in S}$ the OR operation,  $a\wedge b=(a_i b_i)_{i\in S}$ the AND operation and
\begin{eqnarray}
\hat{g}(s,t 
)=
\frac{1}{2}\left(\!\!\begin{array}{c}n \\{n_A}\end{array}\!\!\right)^{\!\!-1}  \left[ \left(\!\!\begin{array}{c}
     n-s-t    \\
      {n_A}-s
\end{array}\!\! \right) +\left(\!\!\begin{array}{c}
     n-s-t    \\
      {n_A}-t
\end{array}\!\! \right)\right] .
\label{eq:hatg}
\end{eqnarray}
Using the definitions we notice the following symmetries of the
coupling function: \barr\label{eq:symmetries}
\left\{\begin{array}{c}
     \Delta(k,k';l,l'
     )=\Delta(k',k;l,l'
     )    \\
      \Delta(k,k';l,l'
      )=\Delta(l,l';k,k'
      )\\
\Delta(k,k';l,l'
)=\Delta(k',k;l',l
)
\end{array} \right. .
\earr

\section{Partition function}
\label{sec:partfun}

In order to study the minimization problem, we will reformulate it
in terms of classical statistical mechanics: in particular, the
minimum $E_0$ of $\pi_{\mathrm{ME}}$ will be recovered in the
zero temperature limit of a suitable classical system.

The main quantity we are interested in is the average bipartite
entanglement between balanced bipartitions, $\pi_{\mathrm{ME}}$ in
Eq.\ (\ref{eq:piave}). This quantity will play the role of energy
in the statistical mechanical approach. We therefore start by
viewing $\pi_{\mathrm{ME}}$ in Eq.\ (\ref{eq:piave}) as a cost
function (potential of multipartite entanglement) and write
\beq
H(z)=\pi_{\mathrm{ME}}(\ket{\psi}),
\eeq
where $z$ are the Fourier coefficients of the expansion
(\ref{eq:genrandomx}). We consider an ensemble $\{m_{j}\}$ of $M$
vectors (states), where $m_j$ is the number of vectors with purity
$H=\epsilon_j$. In the standard ensemble approach to statistical
mechanics one seeks the distribution that maximizes the number of
states $\Omega=M!/\prod_j m_{j}!$ under the constraints that
$\sum_j m_{j} =M$ and $\sum_j m_{j}\epsilon_j =ME$. For $M \to
\infty$, the above optimization problem yields the canonical
ensemble and its partition function
\begin{equation}
\label{partition.function}
\fl\qquad Z(\beta) = \int \d\mu_C(z) \e^{-\beta H(z)}= c_N \int \d\mu_{\mathrm{H}}(U)
\exp\left(-\beta \mathbb{E}[\tr_A (\tr_{\bar{A}} U\ket{\psi_0}\bra{\psi_0}U^\dagger)^2]\right),
\end{equation}
where the expectation value $\mathbb{E}$ was introduced in Eq.\ (\ref{eq:piave}) and the Lagrange multiplier $\beta$, that plays the role of an
inverse temperature, fixes the average value of purity $E$. In the first integral we have used the measure (\ref{eq:meastyp}) and taken into account the normalization condition (\ref{eq:constraint}).
In the last (base-independent) expression $\mu_{\mathrm{H}}$ denotes the Haar measure over
$\mathcal{U}(\mathcal{H})$, $\ket{\psi_0}$ is any given vector and
the (unimportant) constant
$c_N$
is proportional to the ratio $\mu_C(\mathbb{C}^N)/\mu_{\mathrm{H}}(\mathcal{U}(\mathcal{H}))$ between the area of the
$(N-1)$-dimensional sphere (\ref{eq:constraint}) and the volume of
the unitary group.
In conclusion, the potential of multipartite entanglement can be now considered as
the Hamiltonian of a classical statistical mechanical system.

\subsection{Comments}
\label{sec:commpart}

In order to clarify the rationale behind our analysis, a few
comments are necessary.

i) Although our interest is focused on the microcanonical features
of the system, namely on ``isoentangled" manifolds \cite{Kus01}, we find it
convenient to define a canonical ensemble and a temperature. This
makes the analysis easier to handle and is at the very foundations
of statistical mechanics, when one discusses the equivalence in
the description of large systems between the microcanonical
ensemble (in which energy is fixed) and the canonical ensemble (in
which temperature is fixed).

ii) One can view the multipartite system as an ensemble for the
collection of all balanced bipartitions. However, what makes the
problem intricate and interesting is the fact that there is a
nontrivial interaction among different bipartitions, which in
general provokes frustration.

iii) From a physical point of view, the measure  of typical states
is a uniform measure over the whole projective space. This would
be consistent with ergodicity. However, our analysis is purely
static and we are not considering the time evolution generated by
the (purity) Hamiltonian. The relaxation to equilibrium, as well
as its ergodic properties, deserve a deeper study and
would probably uncover additional features with respect to the
equilibrium situation. This aspect will be investigated in the future.

iv) Temperature is a Lagrange multiplier for the optimization
parameter. It is the variable that is naturally conjugate to $H$, in
exactly the same way as inverse temperature is conjugate to energy:
$\beta$ fixes, with an uncertainty that becomes smaller for a larger
system, the level of the purity of the subset of vectors under
consideration, and thus an isoentangled submanifold. The use of a
temperature is a common expedient in problems that can be recast in
terms of classical statistical mechanics. One can find examples of
this kind in the stochastic approach to optimization processes (for
instance simulated annealing) \cite{KGV,MPepl}.

\subsection{Some limits}
\label{sec:limbeta}

We start by looking at some interesting limits and give a few
preliminary remarks. For $\beta\to0$, Eq.\
(\ref{partition.function}) clearly yields the distribution of the
typical states (\ref{eq:meastyp}). For $\beta\to +\infty \; (T\to
0^+)$, only those configurations that minimize the Hamiltonian
survive, namely the MMES. There is a physically appealing
interpretation even for negative temperatures: for $\beta\to -\infty
\; (T\to 0^-)$, those configurations are selected that maximize the
Hamiltonian, that is separable (factorized and non-entangled)
states.

The energy distribution function at arbitrary $\beta$ can be
obtained from the partition function
\barr
\label{eq:partfunc}
Z(\beta)&=&\int \d\mu_C(z) \e^{-\beta H(z)}
=\int_{E_0}^1 \d E \e^{-\beta E }\int \d\mu_C(z)\, \delta(H(z)-E),
\earr
where $E\in[E_0,1]$, $E_0$ being the minimum of the spectrum of $H$ and
$\delta$ the Dirac function.
Incidentally, notice that Eqs.\ (\ref{eq:pmeconstraint}) and (\ref{eq:6a})
yield
\barr
\lim_{n \to \infty} E_0^{(n)} \leq \lim_{n \to \infty} \mu^{(n)} =0, \qquad
E_0^{(n)} \geq 2^{-[n/2]}, \quad \mu^{(n)}=
\frac{2^{[n/2]}+2^{[(n+1)/2]}}{2^n + 1}. \nonumber \\
\earr
The energy distribution function reads
\barr \label{eq:pbeta}
P_\beta(E)&=&\frac{\e^{-\beta E}}{Z(\beta)} \int \d\mu_C(z)\,
\delta(H(z)-E)
\earr
which, for $\beta=0$, simply reads
\barr\label{eq:pzero}
P_0(E)&=&\frac{1}{Z(0)} \int \d\mu_C(z)\, \delta(H(z)-E).
\earr

\begin{figure}
\begin{center}
\includegraphics[width=0.7\textwidth]{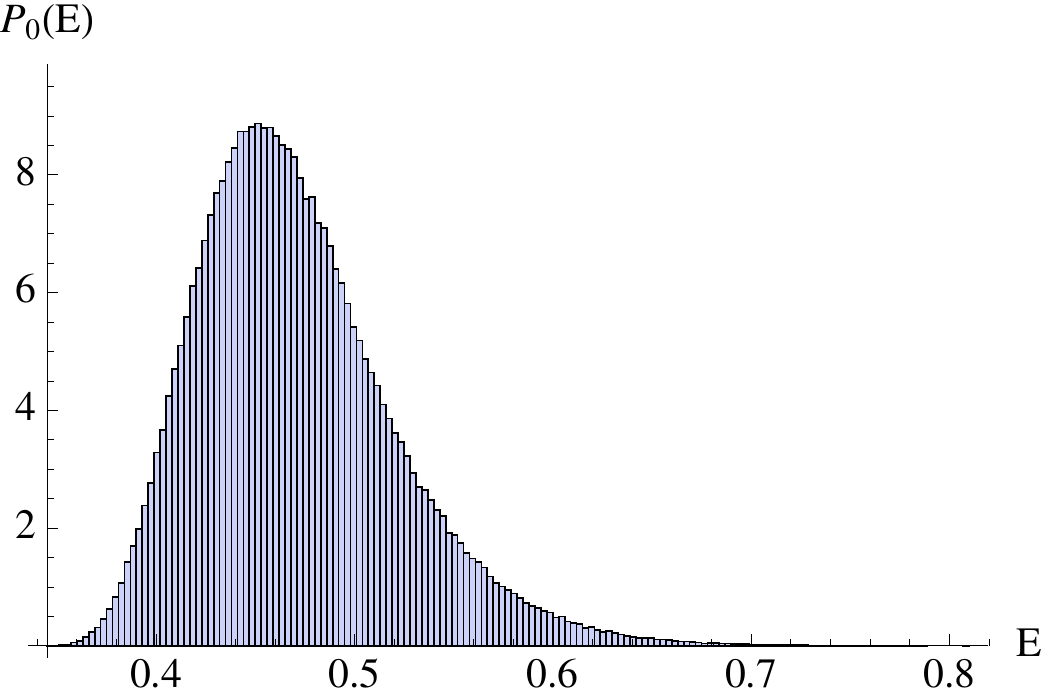}
\caption{(Color online)
Probability density function $P_0(E)$ for $n=4$. The distribution has been obtained from $5\times10^5$ typical states. The binning is $3 \times 10^{-3}$ and the integral is equal to $1$.}
\label{pme4}
\end{center}
\end{figure}

In  Fig.\ \ref{pme4} we show the probability density function $P_0(E)$ for $n=4$. 
As emphasized in Sec.\ \ref{sec:smalln}, this is one of those cases in which frustration appears, as for $n=4$ qubits one (numerically) finds $E_0^{(4)}=\min \pi_{\mathrm{ME}}^{(4)}=1/3 > 1/4 =1 /N_A$ \cite{MMES,sudbery,sudbery2,higuchi}.
We clearly observe the asymmetry of the curve, denoting a positive value of the skewness. This deformation becomes less evident for larger values of $n$. As we will see, in the thermodynamic limit, $n\to\infty$,  $P_0(E)$ will become more and more symmetric and will tend to a Gaussian.

Using Eqs.\ (\ref{eq:partfunc})-(\ref{eq:pzero}) we obtain the expression
of the energy distribution function at arbitrary $\beta$ in terms
of its infinite temperature limit:
\begin{eqnarray}
\label{energy.distribution}
P_\beta(E) =\frac{\e^{-\beta E}P_0(E)}{\int_{E_0}^1 \d E\, \e^{-\beta E}P_0(E)}.
\end{eqnarray}
Notice that this equation is valid at fixed $n$.

By multiplying and dividing the
last equation by $|\beta |\e^{\beta}$ and $\beta \e^{\beta E_0}$,
respectively, and remembering that
\beq
\frac{1}{\epsilon}\,\e^{-x/\epsilon}\stackrel{\epsilon\rightarrow 0}{\longrightarrow} \delta(x)
\eeq
we find
\begin{equation} \label{betalimits}
P_{-\infty}(E)=\delta(E-1), \qquad P_\infty(E)=\delta(E-E_0).
\end{equation}
These limits are the counterparts of those discussed for the
partition function and are reflected in the asymptotic behaviour of
the average energy as function of $\beta$
\begin{eqnarray} \label{mean.energy}
\langle H\rangle_\beta=\frac{1}{Z(\beta)} \int \d\mu_C(z)
H \e^{-\beta H}=\int_{E_0}^1 \d E\, E P_\beta(E)=-\frac{\partial}{\partial\beta}\ln Z(\beta).
\end{eqnarray}
Indeed,
\begin{equation} \label{Hbetalimits}
\langle
H\rangle_{\beta\to-\infty}=1, \quad \langle
H\rangle_{\beta\to+\infty}=E_0.
\end{equation}
More generally, the $m$-th cumulant of $H$ reads
\beq
\kappa^{(m)}_\beta[H] ={(-)}^m \frac{\partial^{m}}{\partial {\beta}^m}
\ln Z(\beta)
={(-)}^{m-1} \frac{\partial^{m-1}}{\partial {\beta}^{m-1}}
\langle H\rangle_\beta.
\eeq
We find
\beq
\label{eq:energyderivative}
\frac{\partial}{\partial\beta}\langle
H\rangle_\beta=-\kappa^{(2)}_\beta[H]=-\langle
H^{2}\rangle_\beta+\langle H\rangle_\beta^{2} \equiv -\bar
\sigma^2_\beta
\leqslant 0,
\eeq
which is non-positive. In particular
\beq
\label{eq:sigma2bar}
\bar \sigma^2=\bar \sigma^2_0=\kappa^{(2)}_0[H].
\eeq
The average energy is a non-increasing function of $\beta$ and has
at least one inflexion point as function of $\beta$. Moreover
\beq
\label{eq:sigmaderivative}
\kappa^{(3)}_\beta[H]=\frac{\partial^2}{\partial\beta^2}\langle
H\rangle_\beta=-\frac{1}{2}\frac{\partial}{\partial\beta}\bar
\sigma^2_\beta.
\eeq
From the qualitative behaviour of $\kappa^{3}_\beta$ one can obtain
information about the width of the distribution. For
$\beta\to+\infty$ the curvature of $\langle H\rangle_\beta$ is
positive and therefore $\bar \sigma ^2_\beta$ is a decreasing
function.

\begin{figure}
\begin{center}
\includegraphics[width=0.98\textwidth]{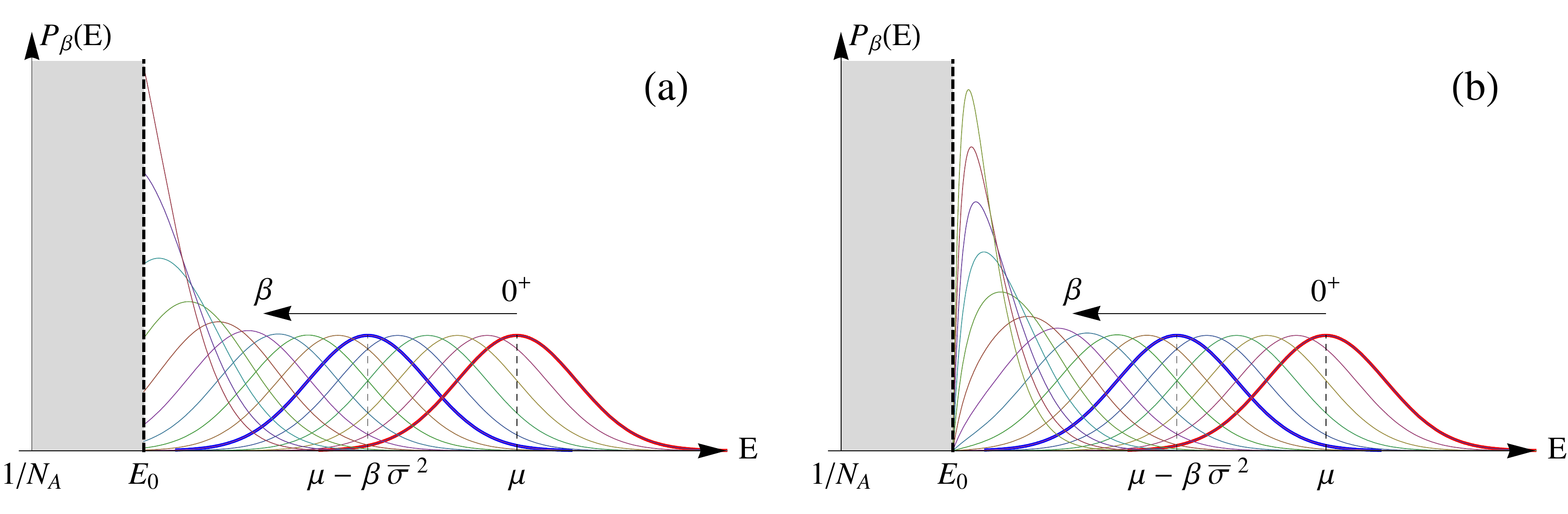}
\caption{(Color online)
Qualitative sketch of Eq.\ (\ref{energy.distribution}), at fixed
$N_A$, in arbitrary units. The energy  density function is
distributed around $\mu$ with standard deviation $\bar{\sigma}$ at
$\beta=0$ (first bell-shaped curve at the right in both panels)
and moves toward $E_0$ when $\beta$ increases. In each panel, from
right to left, $\beta$ changes in constant steps. The probability
density rigidly shifts with $\beta$, for $\beta
\lesssim N^{7/2-\log_2 3}$: see Eq.\ (\ref{eq:shift}) and
following discussion. Note that both $E_0, \mu$ are $O(N^{-1/2})$.
In (a) $P_0(E_0)\neq0$; in (b) $P_0(E_0) = 0$. }
\label{figtdependence}
\end{center}
\end{figure}

From a qualitative point of view, one expects the behavior sketched
in Fig.\ \ref{figtdependence}: for $\beta
\to 0^+ \; (T\to +\infty)$, the distribution is bell-shaped 
(typical states); when $\beta\to +\infty \; (T\to 0^+)$ the
distribution tends to become more concentrated around $E_0$. 
The energy distribution  (\ref{energy.distribution}) at
sufficiently high temperatures [how high will be discussed in Sec.\ \ref{sec:analysis}, see Eq.\ (\ref{eq:shift2})]
can be obtained by observing that from Eq.\ (\ref{eq:energyderivative})

\begin{eqnarray} \label{eq:shiftgen}
P_{\beta}(E)  \sim  P_0(E+\beta \bar{\sigma}^2).
\end{eqnarray}
For larger
values of $\beta$ the left tail of the distribution starts
``feeling" the wall at $E_0$. The value of $P_0(E_0)$ influences the
behaviour of $P_\beta(E)$. 
In general, $P_0(E_0)$ can vanish or not, yielding the behavior
sketched in Fig.\ \ref{figtdependence}(a) and (b), respectively. One finds
\barr\label{eq:betalarge}
P_\beta(E)
\sim \frac{\beta^{r+1}}{r!}(E-E_0)^r \e^{-\beta(E-E_0)} ,
\earr
where $r$ is the order of the first nonvanishing derivative of
$P_0(E)$ at $E_0$. [Figs.\ \ref{figtdependence}(a), (b) display the
case $r=0, 1$, respectively] Notice that the only relic of $P_0(E)$
in (\ref{eq:betalarge}) is $r$ and $P_{\beta \to
\infty}(E)$ yields the second equation in (\ref{betalimits}).
Actually if $r=0$, Eq.\ (\ref{eq:betalarge}) yields a pure
exponential converging to
\beq
P_{+\infty}(E)=\delta(E-E_0).
\eeq
If $r\ge 1$ the probability for finite $\beta$ has an initial
polynomial increase but still converges to a Dirac $\delta$ in
$E_0$, corresponding to MMESs. The analysis for $\beta\to-\infty$ is
analogous (we expand $P_\beta(E)$ around $E=1$, which is the maximum
of $H$); it yields the first equation in (\ref{betalimits}). In this
limit we obtain the separable states.

\section{High temperature expansion}
\label{sec:hight}

This section is devoted to the study of the cumulants of $P_0(E)$.
This will enable us to look at some properties of the high temperature
expansion of the distribution function of the potential of
multipartite entanglement. We remind that for $\beta \to 0$
one gets the typical states.

The high temperature expansion originates from the Taylor series
\barr
\kappa^{(m)}_\beta[H] &=& {(-)}^m
\left. \sum_{j=0}^\infty\frac{\beta^j}{j!}
\left(\frac{\partial^{m+j}}{\partial\beta^{m+j}}\ln Z(\beta)\right)
\right|_{\beta=0} \nonumber \\
&=&\sum_{j=0}^\infty\frac{{(-\beta)}^j}{j!}\kappa^{(m+j)}_{\beta=0}[H].
\label{derivatives}
\earr
The average energy reads
\begin{eqnarray}
\label{high.temp} \langle H\rangle_\beta
&=&
\sum_{m=1}^\infty\frac{(-\beta)^{m-1}}{(m-1)!}\kappa_{\beta=0}^{(m)}[H] \nonumber\\
&\sim&\langle H\rangle_0-\beta \left\langle {\left(H- \langle
H\rangle_0\right)}^2 \right\rangle_0 + \frac{\beta^2}{2}
\left\langle {\left(H- \langle H\rangle_0\right)}^3
\right\rangle_0,
\label{eq:energyhightemp}
\end{eqnarray}
while the free energy takes the form
\barr
F(\beta)&=&\frac{1}{\beta} \ln Z(\beta)\nonumber\\&\sim& \frac{\ln Z(0)}{\beta}-\langle H\rangle_0+\frac{\beta}{2} \left\langle {\left(H- \langle H\rangle_0\right)}^2 \right\rangle_0
-\frac{\beta^2}{6} \left\langle {\left(H- \langle H\rangle_0\right)}^3 \right\rangle_0.
\label{summary}
\earr

In the following three subsections we will evaluate the first three
cumulants of the distribution for $\beta=0$ in order to
characterize the high temperature expansion of the energy
distribution function.

\subsection{First cumulant}

The joint probability density of $z=(z_k)\in\mathbb{C}^N$ associated to the measure
 of typical states (\ref{eq:meastyp}) is
\begin{equation}
p_N(z_1,z_2,\dots,z_N)= \frac{(N-1)! }{\pi^N}\; \delta\!\left(1-\sum_{1\leq k\leq N} |z_k|^2\right).
\label{eq:pN(r)}
\end{equation}
By integrating out $N-M$ variables, one gets
\begin{equation}
p_N(z_1,z_2,\dots,z_M)=
\frac{(N-1)!}{(N-M-1)! \; \pi^M} \,
\left(1-\sum_ {1\leq k\leq M} |z_k|^2\right)^{N-M-1},
\end{equation}
for $1\leq M < N$. In particular the probability density of an
arbitrary element of $z$ is
\begin{equation}
p_N(z_1)=  \frac{N-1}{\pi} \; (1-|z_1|^2)^{N-2}.
\label{eq:pN1}
\end{equation}
Since $\langle e^{i\arg z_j}\rangle_0=0$, the only nonvanishing
averages of the type
\begin{equation}
\left\langle \prod_{k\in X} z_k  \prod_{l\in Y} \bar{z}_l \right\rangle_0 ,
\end{equation}
with $X, Y \subset {\mathbb{Z}}_2^n$, are obtained when the variables
$\{z_{k}\}$ and $\{\bar{z}_{l}\}$ are equal pair by pair, that is
when the sets of indices are equal, $X=Y$. 
The nonvanishing correlation functions are given by
\barr 
\left\langle \prod_{j=1}^{k} |z_{q_j}|^{2m_j} \right\rangle_0
&=& \int \prod_{j=1}^{k} |z_{j}|^{2m_j} p_N (z_1,\dots,z_N) \prod_j
dz_j d\bar{z}_j
\nonumber\\
&=&\frac{(N-1)!\,\prod_{j=1}^{k} m_j!}{\left(N-1+\sum_{j=1}^{k}
m_j\right)!}\;  .
\label{exp}
\earr

A simple proof goes as follows. Extend the product to all $N$ variables by letting some $m_j$ vanish and consider the quantity, with $\alpha_i > 0$,
\begin{eqnarray}
\fl \left\langle \prod_{j=1}^N |z_{j}|^{2 m_j}  \e^{-\sum_k \alpha_k |z_k|^2} \right\rangle_0 & = & 
\int_{(\mathbb{R}^+)^N} \prod_{j}  \left(\d x_j\, x_{j}^{m_j}\right) \e^{-\sum_k \alpha_k x_k} 
(N-1)! \int_{\mathbb{R}} \frac{\d\omega}{2\pi} \e^{-\i\omega(1-\sum_k x_k)}
 \nonumber\\
 & = & (N-1)! \int_{\mathbb{R}} \frac{\d\omega}{2\pi}\, \e^{-\i\omega} \prod_k J_{m_k}(\alpha_k -\i\omega),
\end{eqnarray}
where
\begin{equation}
J_m(z)=\int_{\mathbb{R}^+} x^m \e^{-z x} \d x, \qquad (\Re z >0) .
\end{equation}
Now, we have $J_0(z)=1/z$ and $J_m(z)= (-1)^m \d^m J_0/ \d z= m!/z^{m+1}$, and thus
\begin{eqnarray}
\left\langle \prod_j |z_{j}|^{2 m_j}  \e^{-\sum_k \alpha_k |z_k|^2} \right\rangle_0 & = & 
(N-1)! \int \frac{\d\omega}{2\pi}\, \e^{-\i\omega} \prod_j \frac{(m_j)!}{(\alpha_j-\i\omega)^{m_j+1}} 
\end{eqnarray}
By setting $\alpha_j=\alpha$  for all $j$ we get
\begin{eqnarray}
\fl \left\langle \prod_j |z_{j}|^{2 m_j}  \e^{-\alpha \sum_k  |z_k|^2} \right\rangle_0 & = &
 \e^{-\alpha}  \left\langle \prod_j |z_{j}|^{2 m_j} \right\rangle_0
 \nonumber\\
 & =& 
(N-1)!  \prod_j (m_j)! \int \frac{\d\omega}{2\pi}\, \e^{-\i\omega} \frac{1}{(\alpha-\i\omega)^{N+\sum_j m_j}} ,
\end{eqnarray}
which when $m_j=0$ for all $j$ reads
\begin{equation}
 \e^{-\alpha} =
(N-1)!  \int \frac{\d\omega}{2\pi}\, \e^{-\i\omega} \frac{1}{(\alpha-\i\omega)^{N}} .
\end{equation}
Therefore,
\begin{equation}
 \e^{-\alpha} \left\langle \prod_j |z_{j}|^{2 m_j} \right\rangle_0 = \frac{(N-1)!  \prod_j (m_j)! }{(N+\sum_j m_j -1)}\, \e^{-\alpha}
\end{equation}
 and (\ref{exp}) follows.

The average energy at $\beta=0$ can be easily evaluated and is equal
to the average purity $\mu$ defined in (\ref{eq:6a}):
\begin{equation}
 \langle H\rangle_0= \left\langle \mathbb{E}[ \pi_A ] \right\rangle_0 =
 \mathbb{E}\left[ \langle\pi_A \rangle_0 \right] =
  \langle\pi_A \rangle_0 =
\mu^{(n)} .
\end{equation}
Let us check the above result by direct computation, through (\ref{eq:pimeDelta}). We get
\begin{equation}
\langle H\rangle_0 = \sum_{k,l\in {\mathbb{Z}}_2^{2n}} \Delta (k_1,k_2;l_1,l_2)\langle
z_{k_1}z_{k_2}\bar{z}_{l_1}\bar{z}_{l_2}\rangle_0.
\end{equation}
Now,
\begin{eqnarray}
\langle z_{k_1}z_{k_2}\bar{z}_{l_1}\bar{z}_{l_2}\rangle_0 &=& \langle |z_{1}|^2 |z_{2}|^2 \rangle_0 \left(\delta_{k_1,l_1} \delta_{k_2,l_2} + \delta_{k_1,l_2} \delta_{k_2,l_1} \right)
\nonumber\\
&+&  \left(\langle |z_{1}|^4 \rangle_0- 2 \langle |z_{1}|^2 |z_{2}|^2 \rangle_0\right) \delta_{k_1,l_1} \delta_{k_1,l_2} \delta_{k_1,k_2}
\end{eqnarray}
and thus
\begin{eqnarray}
\langle H\rangle_0 &=& 2 \langle |z_{1}|^2 |z_{2}|^2 \rangle_0 \sum_{k_1,k_2\in {\mathbb{Z}}_2^{n}} \Delta(k_1,k_2;k_1,k_2)  \nonumber\\
&+&   \left(\langle |z_{1}|^4 \rangle_0- 2 \langle |z_{1}|^2 |z_{2}|^2 \rangle_0\right)\sum_{k\in {\mathbb{Z}}_2^{n}} \Delta(k,k;k,k),
\end{eqnarray}
where the symmetry (\ref{eq:symmetries}) was used.

By using (\ref{eq:deltag})
and by setting $k=k_1\oplus k_2$, we get
\begin{equation}
\fl \qquad  \sum_{k_1,k_2} \Delta(k_1,k_2;k_1,k_2) = \sum_{k_1,k_2}  g(0,k_1\oplus k_2) = \sum_{k,k_2}  g(k,0) = N  \sum_{k } g(k,0).
\end{equation}
Since $\delta_{k\wedge 0} = 1 $ and $\sum_{0\leq s \leq n} \delta_{|k|,s}=1$, by using (\ref{eq:gdef}) we can write
\begin{equation}
\fl \qquad \sum_{k } g(k,0) =  \sum_{k} \hat g(|k|,0) =  \sum_{k} \hat g(|k|,0) \sum_{0\leq s \leq n} \delta_{|k|,s} =
 \sum_{0\leq s \leq n}   \hat g(s,0) \sum_{k } \delta_{|k|,s} .
\end{equation}
The number of strings containing $s$ ones is
\begin{equation}
\sum_{k \in {\mathbb{Z}}_2^{n}} \delta_{|k|,s} =
\left(\!\!\begin{array}{c}    n    \\  s  \end{array} \!\!\right),
\end{equation}
and from (\ref{eq:hatg})
\begin{eqnarray}
 \left(\!\!\begin{array}{c}    n    \\  s  \end{array} \!\!\right)   \hat g(s,0) &=& \frac{1}{2}
\left(\!\!\begin{array}{c}    n    \\  s  \end{array} \!\!\right) \left(\!\!\begin{array}{c}n \\{n_A}\end{array}\!\!\right)^{\!\!-1}
\left[
\left(\!\!\begin{array}{c}  n-s    \\ {n_A}-s \end{array}\!\! \right)
+\left(\!\!\begin{array}{c} n-s \\  {n_A} \end{array}\!\! \right)
\right]
\nonumber\\
&=& \frac{1}{2}
\left[
\left(\!\!\begin{array}{c}  n_A    \\ s \end{array}\!\! \right)
+\left(\!\!\begin{array}{c} n_{\bar{A}} \\  s \end{array}\!\! \right)
\right] ,
\label{eq:(ns)g(s0) }
\end{eqnarray}
whence
\begin{equation}
\sum_{k} g(k,0)=\frac{1}{2}
\sum_{s} \left[
\left(\!\!\begin{array}{c}  n_A    \\ s \end{array}\!\! \right)
+\left(\!\!\begin{array}{c} n_{\bar{A}} \\  s \end{array}\!\! \right)
\right] = \frac{1}{2}( N_A + N_{\bar{A}}),
\label{eq:sumgk0}
\end{equation}
where $N_A=2^{n_A}$ and $N_{\bar{A}}=2^{n_{\bar{A}}}$.
Therefore, one gets
\begin{eqnarray}
\sum_{k_1,k_2} \Delta(k_1,k_2;k_1,k_2) =
N\sum_{k } g(k,0)
= \frac{N ( N_A + N_{\bar{A}})}{2}.
\end{eqnarray}
On the other hand,
\begin{equation}
\sum_{k} \Delta(k,k;k,k) = \sum_{k} g (0,0) = \sum_{k} \hat g (0,0) = N,
\end{equation}
because $\hat g (0,0)=1$. Summing up, we get
\begin{equation}
\langle H\rangle_0 = N(N_A+N_{\bar{A}}) \langle |z_{1}|^2 |z_{2}|^2 \rangle_0 +
N \left(\langle |z_{1}|^4 \rangle_0- 2 \langle |z_{1}|^2 |z_{2}|^2
\rangle_0\right),
\end{equation}
and since [see Eq.\ (\ref{exp})]
\begin{equation}
\langle |z_{1}|^2 |z_{2}|^2 \rangle_0 = \frac{1}{2} \langle |z_{1}|^4  \rangle_0 = \frac{1}{N (N+1)},
\end{equation}
we obtain
\begin{equation}
\label{eq:firstcumul}
 \langle H\rangle_0 = \frac{N_A+N_{\bar{A}}}{N+1},
\end{equation}
which equals the value (\ref{eq:6a}) of the average purity $\mu^{(n)}$.
In the thermodynamic limit, $N\to\infty$, with $N_A=N_{\bar{A}}=\sqrt{N}$
\begin{equation}
\label{eq:firstcumul1}
 \langle H\rangle_0 \sim \frac{2}{\sqrt{N}} .
\end{equation}

\subsection{Second cumulant}

The second cumulant is defined as
\beq
\label{eq:definitionk2}
\bar{\sigma}^2=
\kappa_{0}^{(2)}[H] =
\left\langle
H^2\right\rangle_0- \langle H\rangle_0^2.
\eeq
In order to evaluate this quantity we will use a diagrammatic
technique based on the definition of the coupling function $\Delta$
and its properties [Eq.\ (\ref{eq:symmetries})]. We start
considering
\barr
\langle H^2\rangle_0&=&\sum_{k,l\in {\mathbb{Z}}_2^{4n}} \Delta (k_1,k_2;l_1,l_2)\Delta (k_3,k_4;l_3,l_4)\langle
z_{k_1}z_{k_2}z_{k_3}z_{k_4}\bar{z}_{l_1}\bar{z}_{l_2}\bar{z}_{l_3}\bar{z}_{l_4}\rangle_0.
\earr
We must have $\{k_i\}=\{l_j\}$ as sets, that is
\begin{equation}
l_i = k_{p(i)}, \qquad p\in\mathcal{P}_4
\end{equation}
with $1\leq i\leq 4$, where $\mathcal{P}_4$ is the permutation group of $\{1,2,3,4\}$.
Therefore,
\begin{equation}
\fl\langle H^2\rangle_0 = {\sum_{k\in {\mathbb{Z}}_2^{2n}}} \sum_{p\in\mathcal{P}_4} \Delta (k_1,k_2;k_{p(1)},k_{p(2)})\Delta (k_3,k_4;k_{p(3)},k_{p(4)})\langle\langle
|z_{k_1}|^2 |z_{k_2}|^2 |z_{k_3}|^2 |z_{k_4}|^2\rangle\rangle_0,
\end{equation}
where
\begin{equation}
\langle\langle |z_{1}|^{2 m} |z_{2}|^{2 n} |z_{3}|^{2 s} |z_{4}|^{2 t}\rangle\rangle_0 =
\frac{1}{m! p! s! t!} \langle |z_{1}|^{2 m} |z_{2}|^{2 n} |z_{3}|^{2 s} |z_{4}|^{2 t}\rangle_0 .
\end{equation}
The above normalization takes into account the fact that if $k_i=k_j$ for some $i\neq j$ the sum over the permutation group $\mathcal{P}_4$ overcounts the number of different terms. For example, if $k_1=k_2$ and different from the others, we get $m=2$, $n=0$, $s=t=1$, and there is a factor $1/2!$, while, if $k_1=k_2=k_3\neq k_4$, we get  $m=3$, $n=s=0$, $t=1$, and there is a factor $1/3!$.

\begin{figure}
\begin{center}
\includegraphics[width=0.9\textwidth]{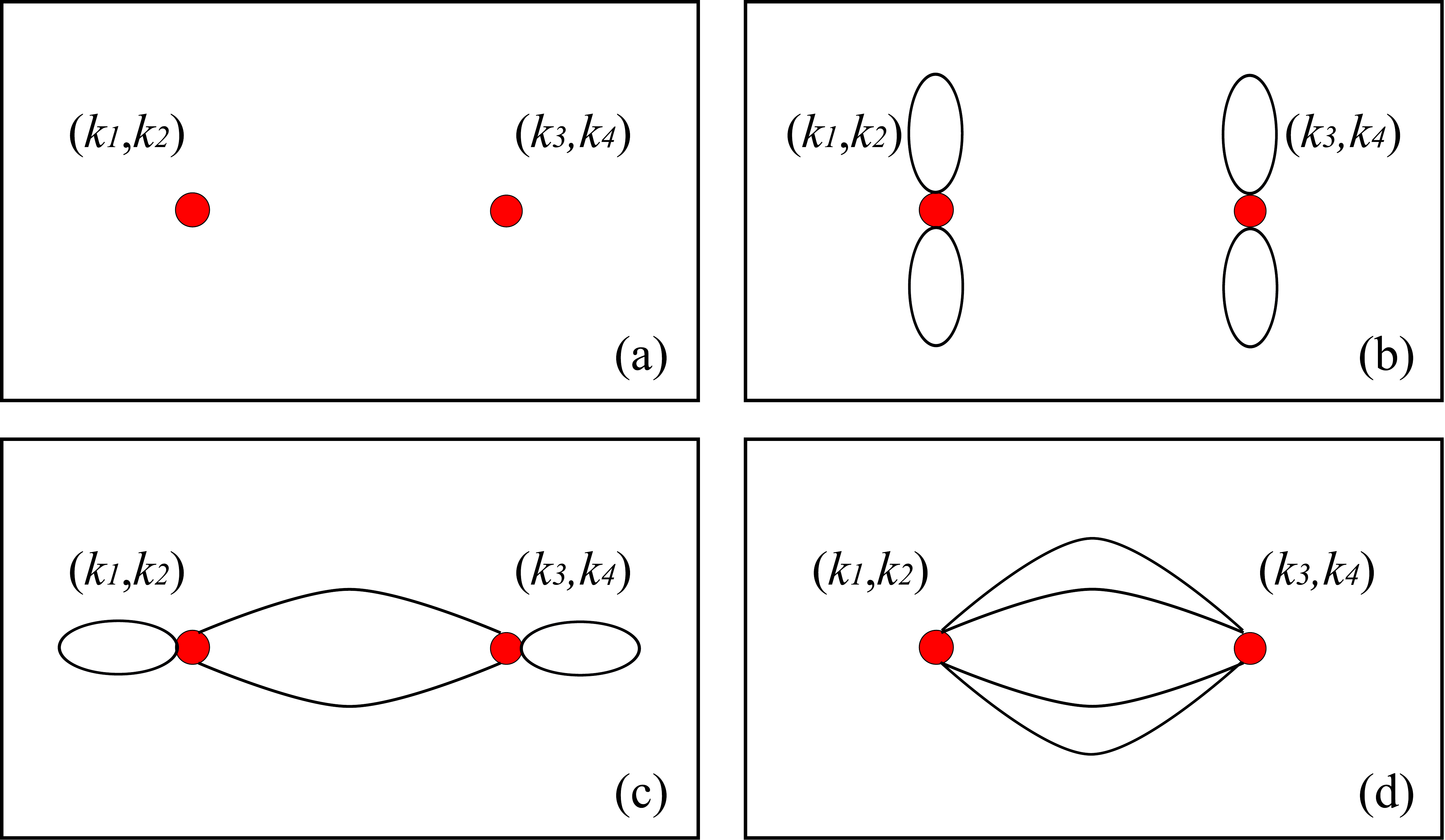}
\caption{(Color online) Graphs used for the evaluation of the second cumulant (\ref{eq:definitionk2}). (a): empty diagram. (b): graph with no links between the left and right pairs. (c): graph with two links between the left and right pairs. (d): graphs with four links between the left and right pairs.}
\label{fig:graphsecondcum}
\end{center}
\end{figure}

Since $m+n+s+t = 4$, from Eq.\ (\ref{exp}) we observe that
\begin{equation}
\langle\langle |z_{1}|^{2 m} |z_{2}|^{2 n} |z_{3}|^{2 s} |z_{4}|^{2 t}\rangle\rangle_0 = \frac{(N-1)!}{(N+3)!} = \frac{1}{N(N+1)(N+2)(N+3)}
\end{equation}
is independent of $k\in {\mathbb{Z}}_2^{2n}$. Therefore,
\begin{equation}
\langle H^2\rangle_0 = \langle |z_{1}|^{2} |z_{2}|^{2} |z_{3}|^{2} |z_{4}|^{2}\rangle_0
\sum_{p\in\mathcal{P}_4}  [p(1)\; p(2), p(3)\; p(4)],
\end{equation}
with  the notation
\begin{equation}
\label{eq:graphequation}
[p(1)\; p(2), p(3)\; p(4)]={\sum_{k\in {\mathbb{Z}}_2^{4n}}} \Delta (k_1,k_2;k_{p(1)},k_{p(2)})\Delta (k_3,k_4;k_{p(3)},k_{p(4)}).
\end{equation}
Note that by the symmetries (\ref{eq:symmetries}) of $\Delta$, we can swap $p(1)\leftrightarrow p(2)$ or $p(3)\leftrightarrow p(4)$, as well as $1\leftrightarrow2$ or $3\leftrightarrow4$, so that
\begin{eqnarray}
\ [w\; x, y\; z ] &=& [x\; w, y\; z ]= [w\; x, z\; y ]=  [x\; w, z\; y ],\nonumber\\
\ [w\; x, y\; z ] &=& [y\; x, w\; z ], \qquad \mathrm{if}\quad w,y \in \{1,2\}, \quad \mathrm{or} \quad w,y \in \{3,4\}.
\label{eq:gsymmetries}
\end{eqnarray}
Using these symmetries we can give a graphical representation of
the quantity in Eq.\ (\ref{eq:graphequation}). Let us consider Fig.\ \ref{fig:graphsecondcum}a. Each vertex
represent a pair $(k_i,k_j)$ in the summation. The edges between vertices and the loops on the same vertex fix the value of $p(i)$. For instance, a double loop on $(k_1,k_2)$ and $(k_3,k_4)$ (see Fig.\ \ref{fig:graphsecondcum}b) yields
\barr
[2\; 1,3\; 4]&=&{\sum_{k}} \Delta (k_1,k_2;k_2,k_1)
\Delta (k_3,k_4;k_{3},k_{4}).
\earr
Each vertex has order 4 with two incoming and two outgoing edges.  Each graph is oriented.
However, for simplicity, in the graphs of Fig.~\ref{fig:graphsecondcum} we have not  indicated the orientations, since in this case, as it is easy to see, they do not yield different contributions. As we shall see, this will not be the case for higher cumulants, where graphs with different orientations represent nonequivalent contributions.

We start considering graphs with no links between the left and
right pairs, see Fig.~\ref{fig:graphsecondcum}b. The sum of this class of
graphs is
\barr
[0-\textrm{link}]=[1\; 2, 3\; 4]+[1\; 2, 4\; 3] + [2\; 1, 3\; 4]+[2\; 1, 4\; 3]=
 4\; [1\; 2, 3\; 4] .
\earr
We have
\begin{eqnarray}
\fl\qquad [1\; 2, 3\; 4]& = & {\sum_k} \Delta(k_1,k_2;k_1,k_2) \Delta(k_3,k_4;k_3,k_4)
= \sum_{k}
g(0, k_1\oplus k_2)\; g(0,k_3\oplus k_4).
\end{eqnarray}
By setting
$l_2= k_1\oplus k_2$ and $l_3= k_1\oplus k_3$,
we get
\begin{eqnarray}
\fl \quad [1\; 2, 3\; 4]= \sum_{k_1,k_4} \sum_{l_2,l_3}
g(0, l_2)\; g(0,l_3)
= N^2 \sum_{l_2,l_3} g(0, l_2) g(0, l_3 )
= N^2 \Big(\sum_{k} g(0, k) \Big)^2  .
\end{eqnarray}
Therefore, by using (\ref{eq:sumgk0}), we get
\begin{equation}
 [0-\textrm{link}]= 4\; [1\; 2, 3\; 4] = N^2 {(N_A+N_{\bar{A}})}^2.
\end{equation}

Let us now consider the graphs with two links between left and right pairs in Fig.~\ref{fig:graphsecondcum}c.
The sum of this class of
graphs is
\begin{eqnarray}
[2-\textrm{link}]&=& [1\; 3, 2\; 4]+[1\; 3, 4\; 2] + [3\; 1, 2\; 4]+[3\; 1, 4\; 2]
\nonumber\\
&+& [2\; 3, 1\; 4]+[2\; 3, 4\; 1] + [3\; 2, 1\; 4]+[3\; 2, 4\; 1]
\nonumber\\
&+& [1\; 4, 2\; 3]+[1\; 4, 3\; 2] + [4\; 1, 2\; 3]+[4\; 1, 3\; 2]
\nonumber\\
&+& [2\; 4, 1\; 3]+[2\; 4, 3\; 1] + [4\; 2, 1\; 3]+[4\; 2, 3\; 1]
\nonumber\\
&=& 16\; [1\; 3, 2\; 4] .
\end{eqnarray}
One gets
\begin{eqnarray}
\fl\quad[1\; 3, 2\; 4] &=& \sum_k \Delta(k_1,k_2; k_1, k_3) \Delta(k_3,k_4; k_2,k_4)
\nonumber\\
&=& \sum_k g\Big( k_2\oplus k_3, (k_1\oplus k_3) \vee  (k_1\oplus k_2)\Big) \;
g\Big(k_2\oplus k_3, (k_2\oplus k_4) \vee (k_3\oplus k_4)\Big).
\nonumber\\
\end{eqnarray}
By setting $l_1=k_1\oplus k_3$, $l_2=k_2\oplus k_3$, and $l_4=k_3\oplus k_4$ we get
\begin{eqnarray}
\fl \qquad  [1\; 3, 2\; 4] &=& \sum_{k_3} \sum_{l_1,l_2,l_4} g\Big( l_2, l_1 \vee  (l_1\oplus l_2)\Big) \;
g\Big(l_2, (l_2\oplus l_4) \vee l_4 \Big)
\nonumber\\
&=& N \sum_{l_1,l_2,l_4} g( l_2, l_1 \vee  l_2) \;
g(l_2, l_2 \vee l_4),
\end{eqnarray}
where we have used the (easy to prove)  useful  relation
\begin{equation}
l_1 \vee (l_1\oplus l_2) = l_1 \vee l_2 .
\label{eq:useful}
\end{equation}
We get
\begin{equation}
l_2\wedge (l_1 \vee  l_2) = (l_1 \wedge l_2) \vee l_2,
\end{equation}
so that the constraint of the function $g$, $l_2\wedge (l_1 \vee  l_2) =0$, implies that $l_2=0$. 
Therefore,  by using (\ref{eq:sumgk0}), we  obtain
\begin{equation}
[2-\textrm{link}]= 16\; [1\; 3, 2\; 4] = 16 N \Big( \sum_k g(k,0) \Big)^2= 4N (N_A+N_{\bar{A}})^2.
\end{equation}

The contribution of the graphs with four links between left and
right pairs (see Fig.~\ref{fig:graphsecondcum}d) has the form
\barr
[4-\textrm{link}]=[3\; 4,1\; 2]+[4\; 3, 2\; 1]+[4\; 3, 1\;2]+[3\;4,2\;1]= 4 [3\; 4,1\; 2].
\earr
We have
\begin{eqnarray}
[3\; 4, 1\; 2]& = & {\sum_k} \Delta(k_1,k_2;k_3,k_4) \Delta(k_3,k_4;k_1,k_2)
=  {\sum_k} \Delta(k_1,k_2;k_3,k_4)^2
\nonumber\\
&=& 
{\sum_k} g\bigg((k_1\oplus k_3) \vee (k_2\oplus k_4) , (k_1\oplus k_4) \vee (k_2 \oplus k_3)\bigg)^2.
\end{eqnarray}
By setting
$l_1= k_1\oplus k_3$, $l_4= k_1\oplus k_4$, and $l_2=k_1\oplus k_2\oplus k_3\oplus k_4$,
we get
\begin{eqnarray}
[3\; 4, 1\; 2] = \sum_{k_3} \sum_{l_1,l_2,l_4}
g(l_1\vee l_2, l_4\vee l_2)^2
= N \sum_{l_1,l_4}
g(l_1, l_4)^2,
\end{eqnarray}
where we used the relation (\ref{eq:useful}) and the constraint $l_2=0$ implied by $(l_1\vee l_2)\wedge(l_4\vee l_2)=0$.
Therefore, we get
\begin{eqnarray}
[4-\textrm{link}]= 4 [3\; 4,1\; 2]
= N f_2(N),
\end{eqnarray}
where
\barr
\label{eq:f2ex}
f_2(N)=4 \sum_{k,l \in {\mathbb{Z}}_2^{n}} g(k,l)^2.
\earr
Notice that if 
\beq
d=|A\cap\bar{B}|=|B\cap\bar{A}|\in[0,n_A].
\eeq
is the distance between bipartitions $(A,\bar{A})$ and $(B,\bar{B})$, defined as the number of qubits belonging to $A$ and not
to $B$, then 
\barr
f_2(N)= 2
\left(\!\!\begin{array}{c} n    \\ n_{A} \end{array}\!\!\right)^{\!\!\!-1}
\!\!\!\sum_{0\leq d\leq n_A}\!\!
\left(\!\!\begin{array}{c} n_A    \\ d \end{array}\!\!\right)
\left(\!\!\begin{array}{c} n_{\bar{A}}    \\ d \end{array}\!\!\right)
2^{n/2}\left[4^{n/4-d}+4^{-(n/4-d)}\right] .
\label{eq:f2appprim}
\earr
See \ref{sec:appendixb}.
Summing up, we obtain
\begin{eqnarray}
\fl\qquad  \sum_{p\in\mathcal{P}_4}  [p(1)\; p(2), p(3)\; p(4)] &=& [0-\textrm{link}] + [2-\textrm{link}] + [4-\textrm{link}]
 \nonumber\\
 &=& 4 [1\; 2,3\; 4]+ 16 [1\; 3,2\; 4]+ 4 [3\; 4,1\; 2]
 \nonumber\\
 &=& N(N+4)(N_A+N_{\bar{A}})^2 + N f_2(N).
\end{eqnarray}
Therefore,
\begin{equation}
\langle H^2\rangle_0 = \frac{f_2(N)+ (N+4)(N_A+N_{\bar{A}})^2}{(N+1)(N+2)(N+3)},
\end{equation}
and
\begin{eqnarray}
\label{eq:finalsigmabar}
\bar{\sigma}^2=\frac{(N+1)f_2(N)-2(N_A+N_{\bar{A}})^2}{(N+1)^2(N+2)(N+3)}.
\end{eqnarray}
We have checked that the above analytic expression of the second cumulant, with $f_2$ given by Eq.\ (\ref{eq:f2ex}), agrees very well (within 1\% up to $n=8$) with the 
 numerical estimates based on the probability density function (obtained by sampling $5\times 10^4$ typical states for each value of $n$).

\begin{figure}
\begin{center}
\includegraphics[width=0.6\textwidth]{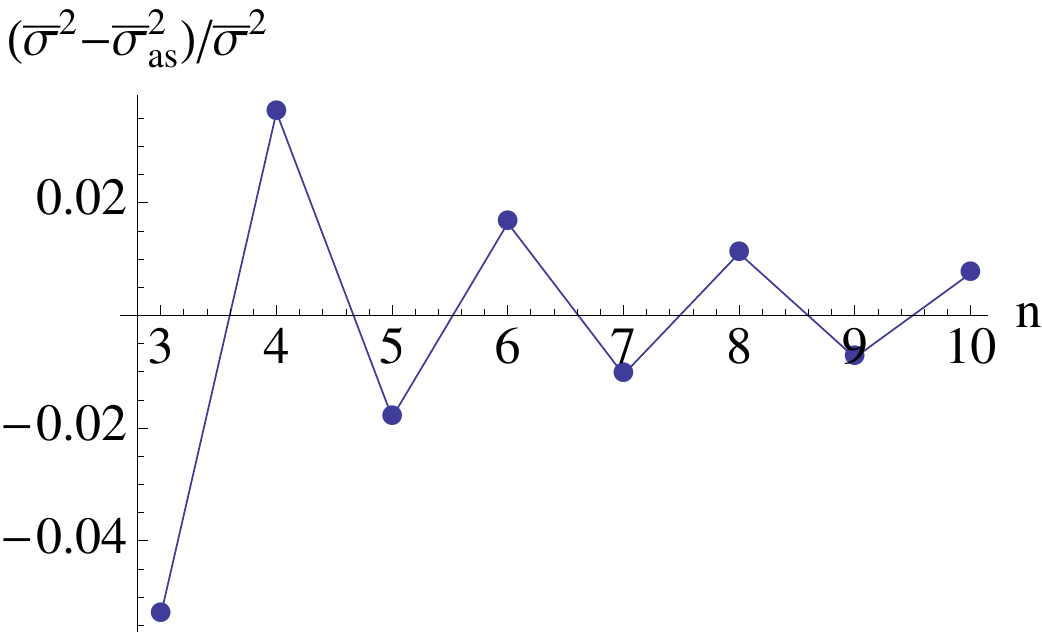}
\caption{(Color online)
Difference between the analytic value of second cumulant, computed
according to Eq.\ (\ref{eq:finalsigmabar}) with $f_2$ given by
(\ref{eq:f2ex}), and its asymptotic limit, obtained by substituting
(\ref{eq:f2}) into Eq.\ (\ref{eq:finalsigmabar}). The approximation
is valid within a few percent even for $n=3$. }
\label{fig:simulvsanal}
\end{center}
\end{figure}

Finally, one proves that (see \ref{sec:appendixb}), in the limit
$N\to\infty$,
\begin{equation}
\label{eq:f2}
f_2(N)\sim 3 \sqrt{2} N^\alpha,
\end{equation}
with
\beq
\alpha=\log_2 3-1 \simeq 0.5850.
\label{eq:alphadef}
\eeq
Therefore, for $N\to\infty$ we
have
\begin{eqnarray} \label{eq:secondocum}
& & \bar{\sigma}^2\sim \frac{f_2(N)}{N^3}=\frac{3 \sqrt{2}}
{N^{4-\log_2 3}} = O(N^{-2.415}).
\end{eqnarray}
Incidentally, note that
\begin{equation}
\bar{\sigma}^2 =\left(\!\!\begin{array}{c}n
\\n_A\end{array}\!\!\right)^{\!\!-2} \sum_{A,B}
\left[ \left\langle \pi_A \pi_B \right\rangle_0 - \left\langle \pi_A  \right\rangle_0\left\langle  \pi_B \right\rangle_0 \right],
\end{equation}
so that, if the bipartitions were independent, we would have
obtained
\begin{equation}
\label{eq:secondocumno}
\bar{\sigma}^2_{\mathrm{ind}} =\left(\!\!\begin{array}{c}n
\\n_A\end{array}\!\!\right)^{\!\!-2} \sum_{A}
\left[ \left\langle \pi_A^2  \right\rangle_0 - \left\langle \pi_A  \right\rangle_0^2 \right]
=\left(\!\!\begin{array}{c}n
\\n_A\end{array}\!\!\right)^{\!\!-1} \sigma^2
\sim N^{-3}.
\end{equation}
Thus, the result in Eq.\ (\ref{eq:secondocum}) \emph{detects an
interference among different bipartitions}. We stress that the
asymptotic estimate is very accurate even for small values of $n$.
In Fig.\ \ref{fig:simulvsanal} we plot the difference between the analytic value of second cumulant, obtained using Eq.\ (\ref{eq:finalsigmabar}) with $f_2$ given by Eq.\ (\ref{eq:f2ex}), and its asymptotic limit, obtained by substituting
(\ref{eq:f2}) into Eq.\ (\ref{eq:finalsigmabar}). We notice an oscillatory behavior: the asymptotic expression systematically overestimates (underestimates) the second cumulant for even (odd) values of $n$. On the other hand, the approximation is very good even for small values of $n$.

\subsection{Third cumulant}
\label{sec:third_cumul}

The third cumulant is defined as
\beq
\label{eq:definitionk3}
\kappa_{0}^{(3)}[H] 
=\left\langle \left(
H- \langle H\rangle_0\right)^3\right\rangle_0=\langle H^3\rangle_0-3\langle H^2\rangle_0\langle H\rangle_0+2\langle H\rangle_0^3.
\eeq
In analogy with the evaluation of the second cumulant we have
\barr
\langle H^3\rangle_0 = \langle |z_{1}|^{2} |z_{2}|^{2} |z_{3}|^{2} |z_{4}|^{2}|z_{5}|^{2} |z_{6}|^{2}\rangle_0
\sum_{p\in\mathcal{P}_6}  [p(1)\; p(2), p(3)\; p(4), p(5)\; p(6)],\nonumber\\
\earr
with 
\barr
\label{eq:graphequation3}
 [p(1)\; p(2), p(3)\; p(4), p(5)\; p(6)]&=&{\sum_{k\in \mathbb{Z}_2^{6n}}} \Delta (k_1,k_2;k_{p(1)},k_{p(2)})\nonumber\\
 &\times&\Delta (k_3,k_4;k_{p(3)},k_{p(4)})\Delta (k_5,k_6;k_{p(5)},k_{p(6)})\nonumber\\
\earr
and
$\mathcal{P}_6$ the permutation group of $\{1,2,3,4,5,6\}$.
From Eq.\ (\ref{exp}) we easily obtain
\barr
\langle |z_{1}|^{2} |z_{2}|^{2} |z_{3}|^{2} |z_{4}|^{2}|z_{5}|^{2} |z_{6}|^{2}\rangle_0=\frac{1}{N(N+1)(N+2)(N+3)(N+4)(N+5)}.\nonumber\\
\earr

\begin{figure}
\begin{center}
\includegraphics[width=0.9\textwidth]{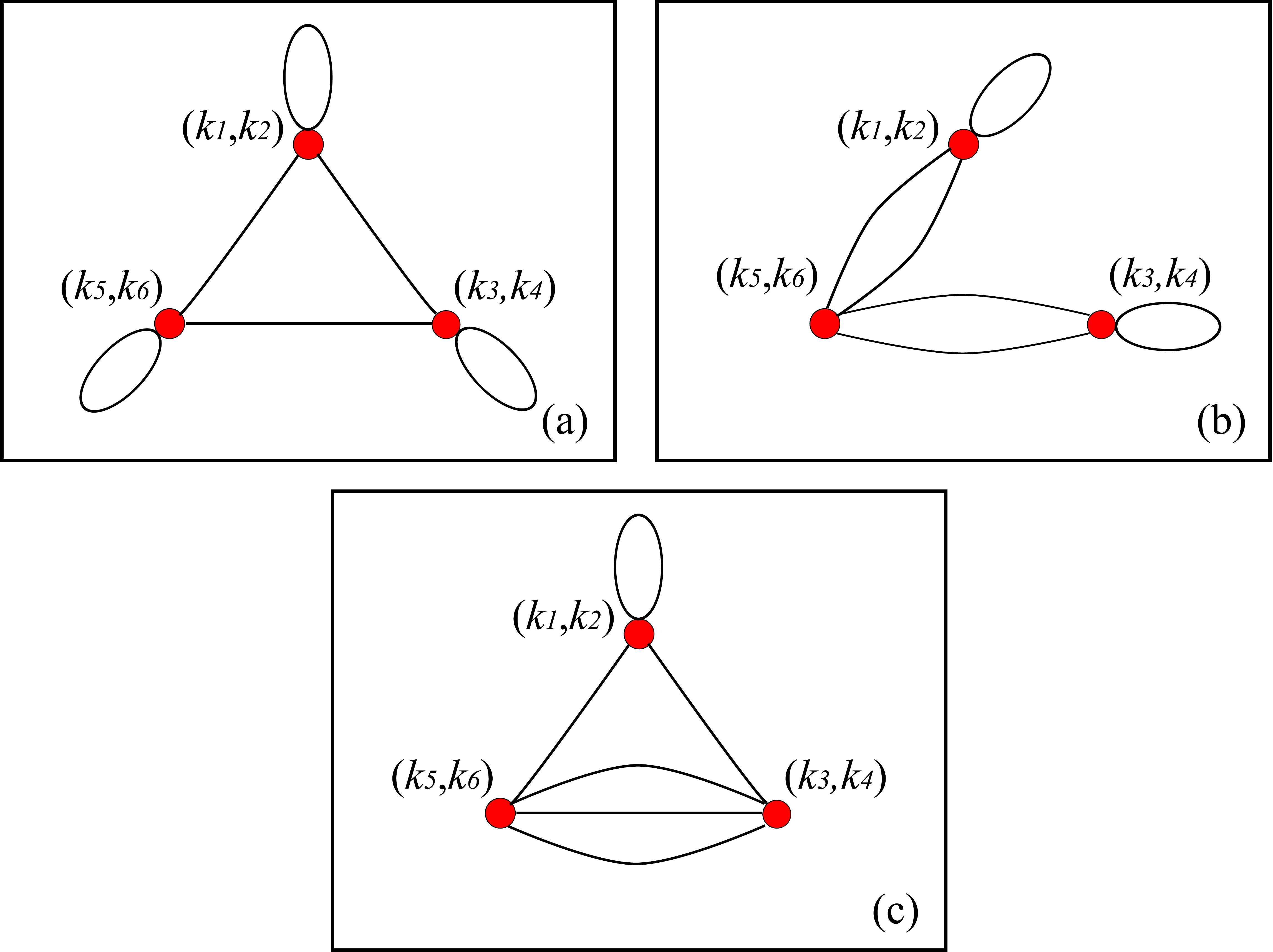}
\caption{(Color online) Non-oriented connected graphs used for the evaluation of the third cumulant. (a): three ears. (b): two ears. (c): one ear. }
\label{fig:graphthirdcumconnected}
\end{center}
\end{figure}

\begin{figure}
\begin{center}
\includegraphics[width=0.9\textwidth]{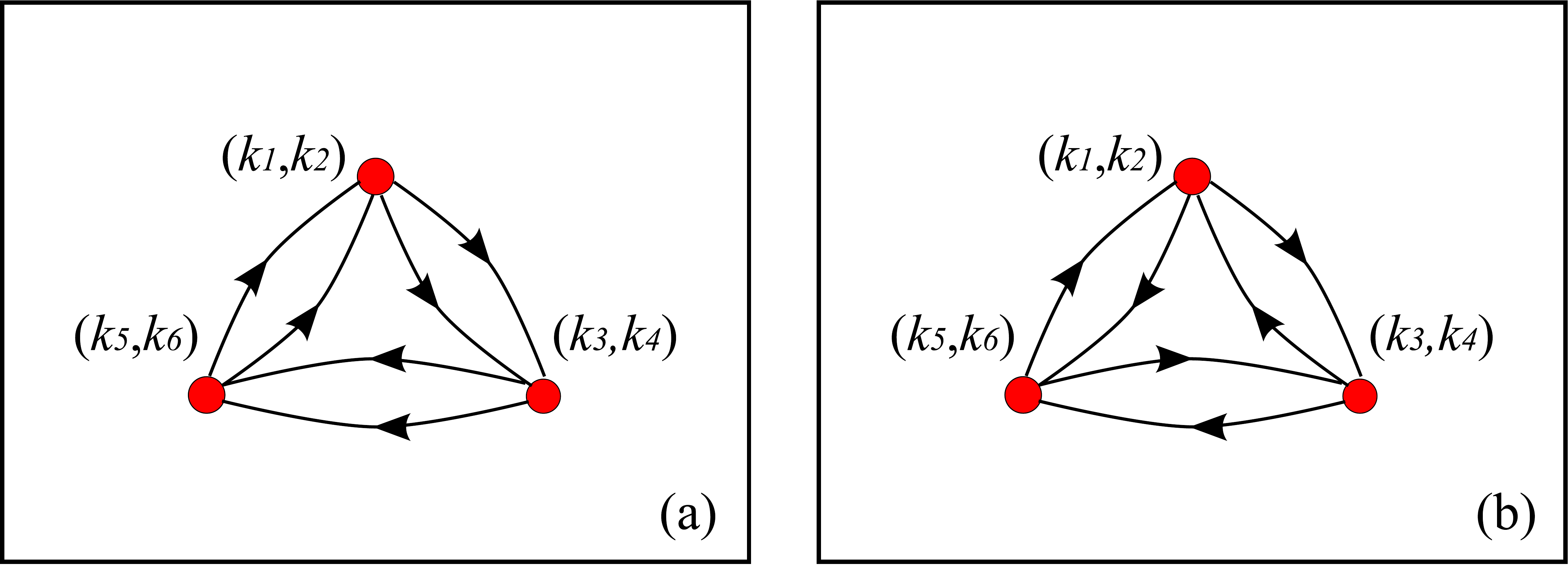}
\caption{(Color online) Oriented connected graphs used for the evaluation of the third cumulant. (a): same internal and external orientations of the edges (nonvanishing ``current").  (b): opposite internal and external orientations of the edges (vanishing ``current"). }
\label{fig:graphthirdcumconnectednoloop}
\end{center}
\end{figure}

\begin{figure}
\begin{center}
\includegraphics[width=0.9\textwidth]{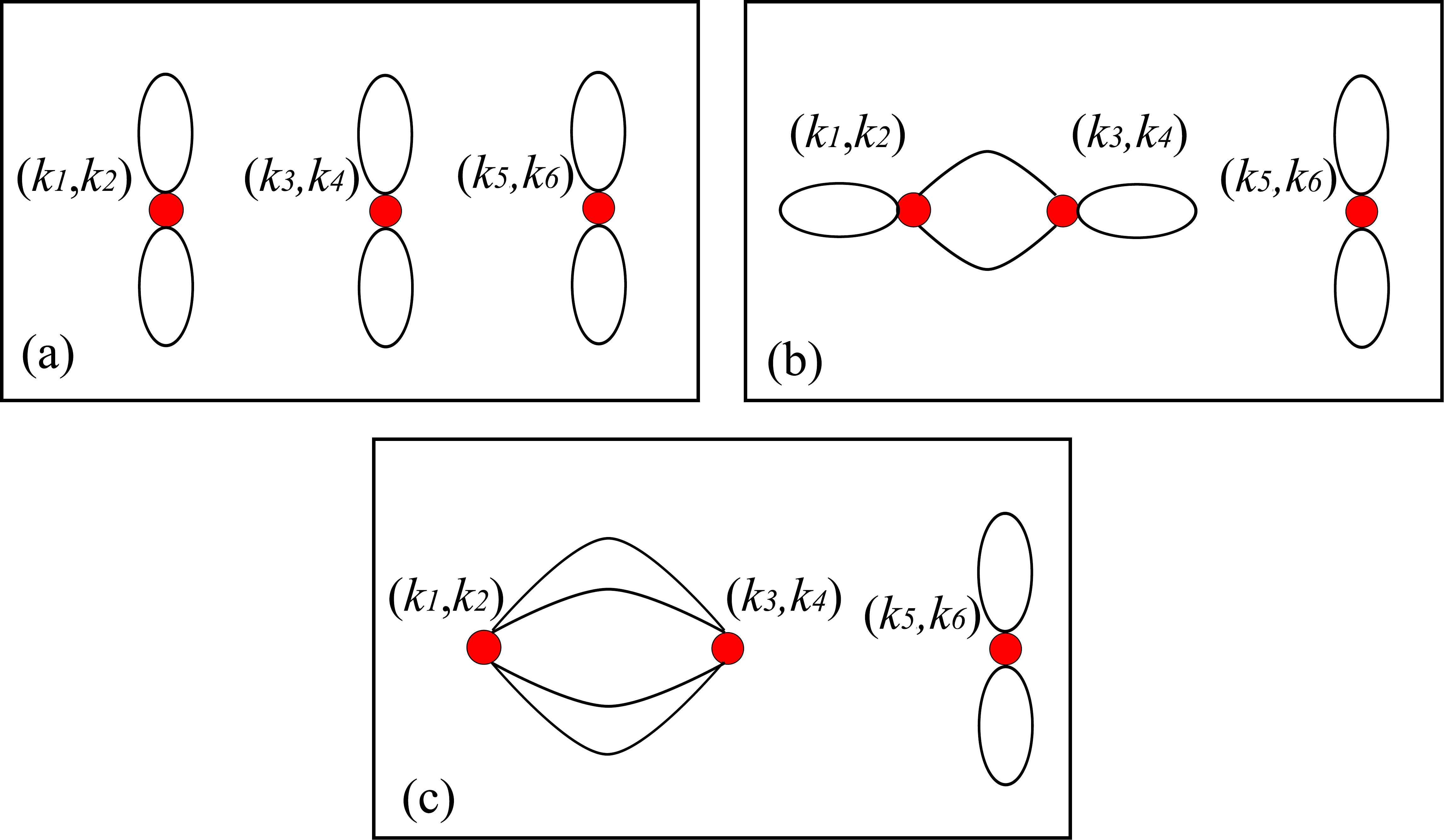}
\caption{(Color online) Disconnected graphs used for the evaluation of the third cumulant. (a): six loops. (b): four loops. (c): two loops.}
\label{fig:graphthirdcumdisconnected}
\end{center}
\end{figure}

We start by considering connected graphs with three ears. A representative of this equivalence class is depicted in Fig.~\ref{fig:graphthirdcumconnected}a. We have
\begin{eqnarray}
\fl\quad[1\; 6, 3\; 2, 5\; 4] &=& \sum_k \Delta(k_1,k_2; k_1, k_6) \Delta(k_3,k_4; k_3,k_2)  \Delta(k_5,k_6; k_5, k_4)
\nonumber\\
&=& \sum_k g\Big( k_2\oplus k_6, (k_1\oplus k_6) \vee  (k_1\oplus k_2)\Big) \;
g\Big(k_2\oplus k_4, (k_2\oplus k_3) \vee (k_3\oplus k_4)\Big)\nonumber\\
&\times&g\Big(k_4\oplus k_6, (k_4\oplus k_5) \vee (k_5\oplus k_6)\Big)\nonumber\\
&=&\sum_{k_1,k_2,k_3,k_5} g(0,k_1\oplus k_2)\;g(0,k_2\oplus k_3)\;g(0,k_2\oplus k_5)\nonumber\\
&=&N \Big(\sum_{k}g(0,k)\Big)^3=N\frac{(N_A+N_{\bar{A}})^3}{8},
\end{eqnarray}
where the constraint in the definition of the function $g$ has implied $k_2=k_4=k_6$ and we used Eq.~(\ref{eq:sumgk0}). The degeneracy of this class of graphs is $128$.

We now consider the class of connected graphs with two ears represented in Fig.~\ref{fig:graphthirdcumconnected}b. We obtain
\begin{eqnarray}
\fl\quad[1\; 3, 2\; 5, 4\; 6] &=& \sum_k \Delta(k_1,k_2; k_1, k_3) \Delta(k_3,k_4; k_2,k_5)  \Delta(k_5,k_6; k_4, k_6)\nonumber\\
&=&\sum_{k_1,k_2,k_4,k_6} g(k_1\oplus k_2,0)\;g(k_2\oplus k_4,0)\;g(k_4\oplus k_6,0)\nonumber\\
&=&N \Big(\sum_{k}g(k,0)\Big)^3=N\frac{(N_A+N_{\bar{A}})^3}{8},
\end{eqnarray}
where we have imposed $k_2=k_3$ and $k_4=k_5$. The degeneracy of the class is $192$.

The final class of non-oriented connected graphs is represented in Fig.~\ref{fig:graphthirdcumconnected}c. Its explicit calculation yields
\begin{eqnarray}
\fl\quad[1\; 6, 2\; 5, 3\; 4] &=& \sum_k \Delta(k_1,k_2; k_1, k_6) \Delta(k_3,k_4; k_2,k_5)  \Delta(k_5,k_6; k_3, k_4)
\nonumber\\
&=& \sum_k g\Big( k_2\oplus k_6, (k_1\oplus k_6) \vee  (k_1\oplus k_2)\Big) \nonumber\\
&\times& g\Big((k_2\oplus k_3)\vee (k_4\oplus k_5), (k_3\oplus k_5) \vee (k_2\oplus k_4)\Big)\nonumber\\
&\times& g\Big((k_3\oplus k_5)\vee (k_4\oplus k_6), (k_4\oplus k_5) \vee (k_3\oplus k_6)\Big)\nonumber\\
&=&\sum_{k_1,...,k_5}g(0,k_1\oplus k_2)g\Big((k_2\oplus k_3)\vee (k_4\oplus k_5), (k_3\oplus k_5) \vee (k_2\oplus k_4)\Big)^2\nonumber\\
&=&\sum_{k_1,...,k_5}g(0,k_1\oplus k_2)g(k_2\oplus k_3, k_2\oplus k_4)^2\nonumber\\
&=&N \sum_k g(0,k) \sum_{l_1,l_2} g(l_1,l_2)^2=N\frac{N_A+N_{\bar{A}}}{2} \sum_{l_1,l_2} g(l_1,l_2)^2\nonumber\\
&=&N\frac{N_A+N_{\bar{A}}}{8} f_2(N),
\end{eqnarray}
where we have used the constraint $k_2=k_6$ and the function $f_2(N)$ defined in Eq.\ (\ref{eq:f2ex}). The degeneracy of this graph is $192$.

In order to take into account the contribution of connected graphs with no ears, it is necessary to consider two different classes of oriented graphs whose representatives are shown in Figs.\ \ref{fig:graphthirdcumconnectednoloop}a and \ref{fig:graphthirdcumconnectednoloop}b, respectively.
For the first class (Fig.\ \ref{fig:graphthirdcumconnectednoloop}a, nonvanishing ``current") we have
\begin{eqnarray}
\fl\quad[5\; 6, 1\; 2, 3\; 4] &=& \sum_k \Delta(k_1,k_2; k_5, k_6) \Delta(k_3,k_4; k_1,k_2)  \Delta(k_5,k_6; k_3, k_4)
\nonumber\\
&=& \sum_k g\Big((k_1\oplus k_5)\vee (k_2\oplus k_6), (k_1\oplus k_6) \vee (k_2\oplus k_5)\Big) \nonumber\\
&\times& g\Big((k_1\oplus k_3)\vee (k_2\oplus k_4), (k_2\oplus k_3) \vee (k_1\oplus k_4)\Big)\nonumber\\
&\times& g\Big((k_3\oplus k_5)\vee (k_4\oplus k_6), (k_4\oplus k_5) \vee (k_3\oplus k_6)\Big)\nonumber\\
&=& \sum_{k_1,k_2,k_3,k_5}  g(k_1\oplus k_5, k_2\oplus k_5) \; g(k_1\oplus k_3, k_2\oplus k_3)\nonumber\\
&\times& g(k_3\oplus k_5, k_1\oplus k_2\oplus k_3\oplus k_5)\nonumber\\
&=&N\sum_{k_1,k_2,k_3}g(k_1,k_2) \; g(k_1\oplus k_3, k_2\oplus k_3)\; g(k_3,k_1\oplus k_2\oplus k_3)\nonumber\\
&=&N\sum_{k_1,k_2,k_3}g (k_1,k_2\oplus k_3 ) \; g (k_2, k_1\oplus k_3 )\; g(k_3,k_1\oplus k_2)\nonumber\\
&=&N \, f_3^{(1)}(N)
\end{eqnarray}
where we have used the constraints $k_4=k_1\oplus k_2 \oplus k_3$ and $k_6=k_1\oplus k_2 \oplus k_5$ and defined
\beq\label{eq:f31}
f_3^{(1)}(N)=\sum_{k_1,k_2,k_3}g(k_1,k_2\oplus k_3) \; g(k_2, k_1\oplus k_3)\; g(k_3,k_1\oplus k_2).
\eeq
The degeneracy of this graph is $16$.
An analogous calculation can be carried out for the second class of oriented graphs 
(Fig.\ \ref{fig:graphthirdcumconnectednoloop}b, vanishing ``current"). We obtain
\begin{eqnarray}
\fl\quad[3\; 6, 5\; 2, 1\; 4] &=& \sum_k \Delta(k_1,k_2; k_3, k_6) \Delta(k_3,k_4; k_5,k_2)  \Delta(k_5,k_6; k_1, k_4)
\nonumber\\
&=&N\sum_{k_1,k_2,k_3}g(k_1,k_2) \; g(k_3,  k_2)\; g(k_1\oplus k_3, k_2)=N \, f_3^{(0)}(N)
\end{eqnarray}
with
\beq\label{eq:f30}
f_3^{(0)}(N)=\sum_{k_1,k_2,k_3}g(k_1,k_2) \; g(k_3,  k_2)\; g(k_1\oplus k_3, k_2). 
\eeq
In this case, the degeneracy is $64$.

The contribution of disconnected graphs (Fig.\ \ref{fig:graphthirdcumdisconnected}) can be computed by considering the results obtained for the first and second cumulant.  
For the class of graphs represented in Fig.\ \ref{fig:graphthirdcumdisconnected}a we have
\begin{eqnarray}
\fl\quad[1\; 2, 3\; 4, 5\; 6] &=& N^3\frac{(N_A+N_{\bar{A}})^3}{8} ,
\end{eqnarray}
with degeneracy $8$.
In the case of the graph in Fig.\ \ref{fig:graphthirdcumdisconnected}b the result is
\begin{eqnarray}
\fl\quad[1\; 3, 2\; 4, 5\; 6] &=& N\frac{(N_A+N_{\bar{A}})^2}{4} N \frac{(N_A+N_{\bar{A}})}{2}= N^2\frac{(N_A+N_{\bar{A}})^3}{8},
\end{eqnarray}
with degeneracy $96$.
Finally from the disconnected graphs with two loops (Fig.\ \ref{fig:graphthirdcumdisconnected}c) we obtain
\begin{eqnarray}
\fl\quad[3\; 4, 1\; 2, 5\; 6] &=&N^2\frac{(N_A+N_{\bar{A}})}{8}f_2(N) ,
\end{eqnarray}
with degeneracy $24$.
In conclusion, we find
\barr
\langle H^3\rangle_0 &=& \frac{1}{(N+1)(N+2)(N+3)(N+4)(N+5)}\nonumber\\
& &\times \Big(16f_3^{(1)}(N)+64f_3^{(0)}(N)+ 3 (N+8)(N_A+N_{\bar{A}})f_2(N)\nonumber\\
& & \qquad + (N_A+N_{\bar{A}})^3(N^2+12N+40)\Big)
\earr
and, therefore,
\begin{eqnarray}
\label{eq:finalk3}
\fl\qquad \kappa_{0}^{(3)}[H] 
&=& \frac{1}{(N+1)^3(N+2)(N+3)(N+4)(N+5)}\nonumber\\
\fl\qquad & & \times \Big(16(N+1)^2 f_3^{(1)}(N)+64(N+1)^2f_3^{(0)}(N)\nonumber\\
\fl\qquad & & \qquad - 36 (N+1)(N_A+N_{\bar{A}})f_2(N)
- 8(N_A+N_{\bar{A}})^3(N-5)\Big).
\end{eqnarray}
We have checked that the above analytic expression of the third cumulant, with $f_2$, $f_3^{(0)}$ and $f_3^{(1)}$ given by
(\ref{eq:f2ex}), (\ref{eq:f30}) and (\ref{eq:f31}), respectively, agrees very well (less than 1\% for $n=1\div7$ and a few \% for $n=8$) with the 
numerical estimates based on the probability density function (obtained by sampling $5\times 10^4$ typical states for each value of $n$).

Finally, in the limit $N\to\infty$, one can prove that (see \ref{sec:appendixe})
\begin{equation}
\label{eq:f301}
f_3^{(0)}(N)\sim c\, N^{5-\gamma},
\end{equation}
with $c\simeq  1.05385$ and $\gamma\simeq 4.1583$ given by (\ref{eq:cdef1}) and (\ref{eq:gammadef1}),
and that
(see \ref{sec:appendixe2})
\begin{equation}
\label{eq:f311}
f_3^{(1)}(N) \sim N^\alpha,
\end{equation}
with $\alpha\simeq 0.5850$ given by (\ref{eq:alphadef}).
Therefore, by recalling the asymptotic expression for $f_2(N)$  (\ref{eq:f2}),
we find that the graph in Fig.\ \ref{fig:graphthirdcumconnectednoloop}b dominates over that in Fig.\ \ref{fig:graphthirdcumconnectednoloop}a and
\begin{eqnarray}
\label{eq:finalk3Nexact}
\kappa_{0}^{(3)}[H] 
&\sim& 64\, c\, N^{-\gamma} \simeq
 67.443 \, N^{-4.1583}.
 \end{eqnarray}

\begin{figure}
\begin{center}
\includegraphics[width=0.6\textwidth]{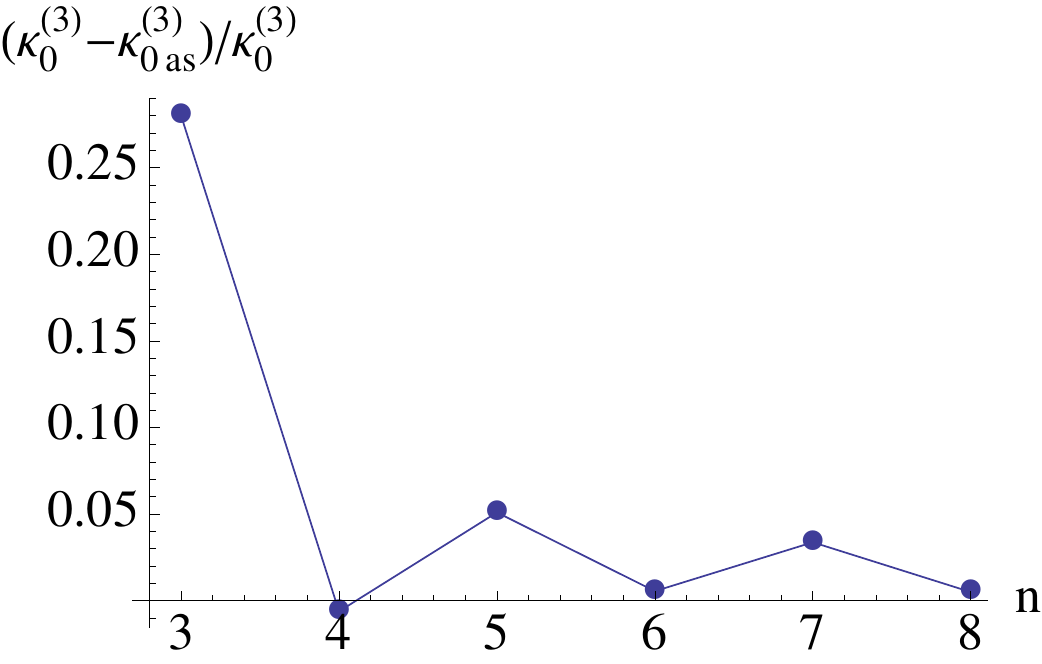}
\caption{(Color online)
Difference between the analytic value of third cumulant, computed
according to Eq.\ (\ref{eq:finalk3}) with $f_2$, $f_3^{(0)}$ and $f_3^{(1)}$ given by
(\ref{eq:f2ex}), (\ref{eq:f30}), (\ref{eq:f30}), and their asymptotic limits, obtained by substituting
(\ref{eq:f2}),  (\ref{eq:f301}),  (\ref{eq:f311}) into Eq.\ (\ref{eq:finalk3}). The approximation
is valid within a few percent for $n\ge 4$. }
\label{fig:simulvsanalcum3}
\end{center}
\end{figure}

In Fig.\ \ref{fig:simulvsanalcum3} we plot the difference between the analytic value of third cumulant  and its asymptotic limit, obtained by substituting (\ref{eq:f2}),  (\ref{eq:f301}),  (\ref{eq:f311}) into Eq.\ (\ref{eq:finalk3}). We again observe an oscillating behaviour. The approximation is good for $n\ge 4$.

\subsection{Gaussian approximation}
\label{sec:analysis}

We can now summarize the results obtained for the first three cumulants and try to get a broader picture.
Equations 
(\ref{eq:firstcumul}), 
(\ref{eq:finalsigmabar}) and
(\ref{eq:finalk3}) are all exact. Their asymptotic expansions for large $N$, are given in Eqs.\ 
(\ref{eq:firstcumul1}),
(\ref{eq:secondocum}) and
(\ref{eq:finalk3Nexact}). 
By plugging these results into Eqs.\ (\ref{high.temp}) and (\ref{summary}) we obtain
the asymptotic expressions of the average energy
\begin{eqnarray}
\label{high.tempexp} \langle H\rangle_\beta
&\sim &
\frac{2}{\sqrt{N}}
-\beta \frac{3 \sqrt{2}}
{N^{3-\alpha}} + \frac{\beta^2}{2}
\frac{64 c}{N^{\gamma}} \nonumber \\
& \simeq &  \frac{2}{N^{0.5}}
-\beta \frac{4.243}
{N^{2.415}} + \frac{\beta^2}{2} \frac{67.443}{N^{4.1583}}
\label{eq:energyhightemp1}
\end{eqnarray}
and the free energy 
\barr
F(\beta)&\sim&
 \frac{\ln Z(0)}{\beta}-\frac{2}{\sqrt{N}}+\frac{\beta}{2} \frac{3 \sqrt{2}}
{N^{3-\alpha}}
-\frac{\beta^2}{6} \frac{64c}{N^{\gamma}}\nonumber\\
&\simeq&\frac{\ln Z(0)}{\beta}-\frac{2}{N^{0.5}}+\frac{\beta}{2} \frac{4.243}
{N^{2.415}}
-\frac{\beta^2}{6} \frac{67.443}{N^{4.1583}}
\label{summaryexp}
\earr
where 
\beq
c\simeq1.054, \quad \alpha=\log_2 3 -1 \simeq 0.5850, \quad \gamma=4.1583.
\eeq
See Eqs.\ 
(\ref{eq:cdef1}) and
(\ref{eq:gammadef1}).

If $N$ is large enough and the first two cumulants at $\beta=0$ suffice, the energy distribution  (\ref{eq:pzero}) can be taken to be Gaussian
\begin{equation} \label{gauss.approx}
P_{0}(E)\sim \frac{1}{\sqrt{2\pi\bar{\sigma}^2}}\exp\left(-\frac{\left(E-\mu\right)^{2}}{2\bar{\sigma}^2}\right),
\end{equation}
where $\mu$ and $\bar{\sigma}$ are given in (\ref{eq:firstcumul1})
and (\ref{eq:secondocum}), respectively. The energy distribution at
arbitrary temperature is then [see Eq.\ (\ref{eq:shiftgen})]
\begin{eqnarray} 
\label{eq:shift}
P_{\beta}(E)  \sim  \frac{1}{\sqrt{2\pi\bar{\sigma}^2 }}
\exp\left(-\frac{\left(E-\mu+\beta \bar{\sigma}^2 \right)^{2}}
{2 \bar{\sigma}^2}\right) .
\end{eqnarray}
This is valid for relatively small $\beta$:
\beq
\label{eq:shift2}
\mu-\beta\bar{\sigma}^2-\bar{\sigma} \gtrsim 0
\quad \Leftrightarrow \quad 
\beta \lesssim \mu/\bar{\sigma}^2 \sim N^{7/2-\log_2 3}.
\eeq
Up to this value the probability density rigidly shifts with
$\beta$, as is apparent in Fig.\ \ref{figtdependence}, 
which was  obtained by numerically solving Eq.\ (\ref{energy.distribution}).

\subsection{A few comments}
\label{sec:comments}

The behavior of the cumulants derived in this section is very peculiar. 
The second and third cumulant follow a nontrivial power dependence, with trascendental exponents [see Eqs.\ (\ref{high.tempexp})]).
Interestingly, close scrutiny of the calculation in Sec.\ 
\ref{sec:third_cumul} shows also that $3-\alpha$, the exponent that governs the $N$-dependence of $\bar{\sigma}^2$, is found in a class of (nondominant) graphs that appear in the evaluation of the third cumulant: the exponent $5-\alpha$ stems from the graph in 
Fig.\ \ref{fig:graphthirdcumconnectednoloop}a (the dominant exponent $\gamma$ stemming from the graph in
Fig.\ \ref{fig:graphthirdcumconnectednoloop}b).
This might suggest a possible recursion of the exponent $\alpha$ at all orders in the cumulant expansion. At this stage, we are unable to say if at higher orders the dominant graph for $\kappa_0^{(3)}$ in 
Fig.\ \ref{fig:graphthirdcumconnectednoloop}b cancels, 
yielding a series in $N^{q(\alpha)}$ with $q$ a function of $\alpha$.

It would be important to go beyond the Gaussian approximation in order to evaluate the behaviour of the left tail of the probability density function, close to $\pi_{\mathrm{ME}} = E \simeq E_0$. See Figs.\ 
\ref{pme4} and \ref{figtdependence}.
This would give us some precious information about the features of MMES and the very structure of entanglement frustration \cite{frustr}. In particular, it would be interesting to understand the role played by the interference among the bipartitions, in connection with the appearance of frustration in MMESs. 
See for instance the asymptotic behavior of the second cumulant  in Eqs.\ (\ref{eq:secondocum})-(\ref{eq:secondocumno}) and the short
discussion that follows.
Additional investigation is necessary in order to elucidate these intriguing issues.

\section{Concluding remarks and outlook}
\label{sec:conclusion}
\label{sec:concl}

We have built a statistical mechanical approach to multipartite
entanglement, by introducing a partition function in order to tackle
a complex optimization problem, whose solutions are the maximally
multipartite entangled states, that appear as minimal energy
configurations.

The scheme adopted here is general. In classical statistical
mechanics, temperature is used to fix the energy to a given value in
the thermodynamic limit. Analogously, the fictitious temperature
introduced here localizes the measure on a set of states whose
entanglement (energy) is fixed, and can be larger or smaller than the
entanglement associated to typical states.

Remarkably, a strategy like the one adopted in this article, when
applied to the simpler case of bipartite entanglement (at a fixed
bipartition) \cite{onebipartition} brings to light an involved landscape of phase
transitions for the purity. Clearly, the multipartite version of the
problem is much more involved, as the picture that emerges is
complex and unearths a remarkable interplay between multipartite
entanglement and frustration. It would therefore be of great
interest to understand whether the phase transition that occurs in
the bipartite situation, when there is no average over the
bipartitions, survives and has a counterpart in the multipartite
scenario. This possibility will be explored in the future.

One important property that we have not investigated here and that
is often used to characterize multipartite entanglement is the
so-called monogamy of entanglement \cite{multipart1,KDS}, that
essentially states that  entanglement cannot be freely shared among
the parties. Interestingly, although monogamy is a typical property
of multipartite entanglement, it is expressed in terms of a bound on
a sum of \emph{bipartite} entanglement measures. This is reminiscent
of the approach taken in this paper. The curious fact that bipartite
sharing of entanglement is bounded might have interesting
consequences in the present context. It would be worth understanding
whether monogamy of entanglement generates frustration.

Finally, we think that the characterization of multipartite entanglement
proposed here can be important
for the analysis of the entanglement features of
many-body systems, such as spin systems and systems close to
criticality.

\ack This work is partly supported by the European Union
through the Integrated Project EuroSQIP.


\appendix

\section{}
\label{sec:appendixa}

\begin{figure}
\begin{center}
\includegraphics[width=0.4\textwidth]{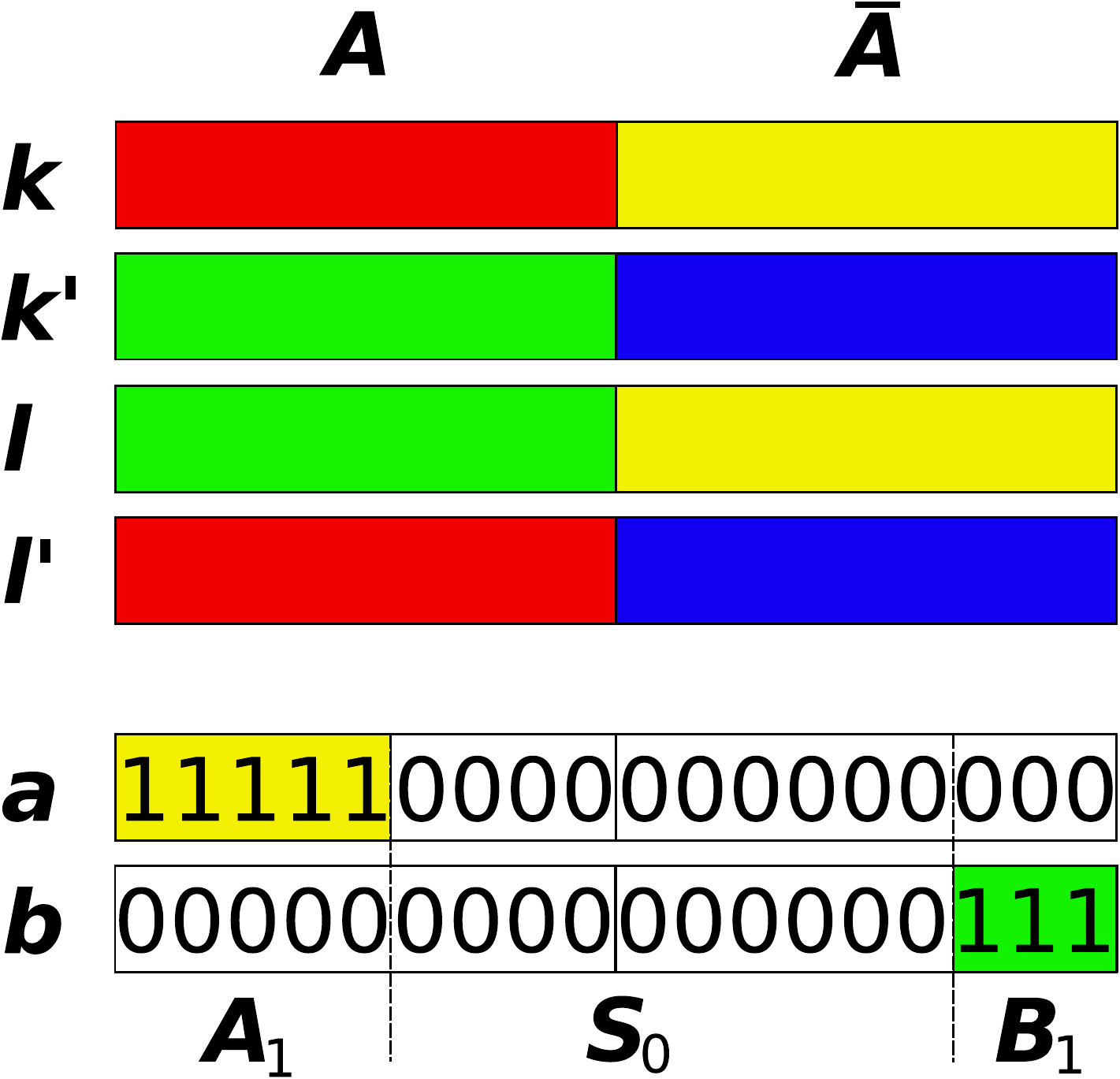}
\caption{(Color online)
Graphic representation of the combination of binary strings contributing to sum (\ref{eq:Deltadef1}).}
\label{figkkll}
\end{center}
\end{figure}

We derive here the expression (\ref{eq:deltag}) of the coupling function. See \cite{multent}.
We start from the definition (\ref{eq:Deltadef}), that can be rewritten as
\begin{equation}
\Delta(k,k';l,l')
=\frac{1}{2} \left(\!\!\begin{array}{c}n \\n_A\end{array}\!\!\right)^{\!\!-1} \left(\tilde\Delta(k,k';l,l';[n/2]) + \tilde\Delta(k',k;l,l';[n/2])\right),
\label{eq:DeltadefA}
\end{equation}
where
\begin{eqnarray}
\tilde \Delta(k,k';l,l';n_A)
= \sum_{|A|=n_A} \delta_{k_{A}, l'_{A}} \delta_{k'_{A},l_{A}}
\delta_{k_{\bar{A}}, l_{\bar{A}}} \delta_{k'_{\bar{A}}, l'_{\bar{A}}} .
\label{eq:Deltadef1}
\end{eqnarray}
Let us fix a quadruple of binary strings $(k,k',l,l')$ and a dimension $n_A$. See figure \ref{figkkll}. A bipartition $(A,\bar{A})$, with $|A|=n_A$ yields a nonvanishing contribution to the sum (\ref{eq:Deltadef1}) when
\begin{equation}
\delta_{k_{A}, l'_{A}}
\delta_{k'_{A},l_{A}}
 \delta_{k_{\bar{A}}, l_{\bar{A}}} \delta_{k'_{\bar{A}},
l'_{\bar{A}}} =1 ,
\end{equation}
that is when
\begin{equation}
k_A=l'_A,\quad k'_A=l_A,\quad k_{\bar{A}}=l_{\bar{A}},\quad k'_{\bar{A}}=l'_{\bar{A}},
\end{equation}
where we recall that $k_A=l_A$ means that the substrings of $k$ and $l$ are equal, namely $k_i=l_i$ for all $i\in A$. By noting that two bits $k_i$ and $l_i$ are equal when $k_i\oplus l_i =0$,  the above condition can be rephrased as
\begin{equation}
k_A\oplus l'_A=0 ,\quad k'_A\oplus l_A=0,\quad k_{\bar{A}}\oplus l_{\bar{A}}=0,\quad k'_{\bar{A}}\oplus l'_{\bar{A}}=0,
\end{equation}
that is
\begin{equation}
(k_A\oplus l'_A) \vee (k'_A\oplus l_A) =0,\quad (k_{\bar{A}}\oplus l_{\bar{A}})\vee (k'_{\bar{A}}\oplus l'_{\bar{A}})=0.
\end{equation}
Summarizing, a bipartition $(A,\bar{A})$ yields a nonvanishing contribution to (\ref{eq:Deltadef}) if and only if
the following substrings are zero
\begin{equation}
a_{\bar{A}} = 0 \quad \mathrm{and} \quad b_A =0 ,
\label{eq:aAbB}
\end{equation}
where
\begin{equation}
a=(k\oplus l) \vee (k' \oplus l') \quad
\mathrm{and}\quad  b=(k \oplus l') \vee (k'\oplus l).
\end{equation}
Note that equation (\ref{eq:aAbB}) implies that
\begin{equation}
\label{eq:awb0}
a\wedge b =0,
\end{equation} since $(a\wedge b)_A = a_A \wedge 0 =0$ and $(a\wedge b)_{\bar{A}} = 0 \wedge b_{\bar{A}} =0$. On the other hand, the substrings $a_A$ and $b_{\bar{A}}$ are totally free, whence
\begin{equation}
\label{eq:awb1}
|a|=|a_A|\leq |A| =  n_A, \qquad
|b|=|b_{\bar{A}}|\leq |\bar{A}| =  n_{\bar{A}}. \qquad
\end{equation}
It is easy to see that (\ref{eq:awb0}) and (\ref{eq:awb1}) are also sufficient conditions for the existence of a partition $(A,\bar{A})$ that satisfies (\ref{eq:aAbB}).

In conclusion, $\tilde\Delta(k,k';l,l';n_A)\neq 0$ when
\begin{equation}
\label{eq:awb2}
a\wedge b=0, \quad \mbox{ with } \quad  |a|\leq n_A, \, |b|\leq
n_{\bar{A}}.
\end{equation}
Therefore,
\begin{equation}
\label{eq:deltaappendix}
\fl \qquad
\tilde\Delta(k,k';l,l';n_A)=\delta_{a\wedge b,\,0}\; \chi_{[0,n_A]}(|a|)\;  \chi_{[0,n_{\bar{A}}]}(|b|)\; \#(k,k',l,l'),
\end{equation}
where $\chi_G$ is the characteristic function of set $G$ and
$\#(k,k',l,l')$ the number of terms
in the sum in (\ref{eq:DeltadefA}) that contribute to the function
$\tilde\Delta$.

Note that, since by (\ref{eq:awb2}) the strings $a$ and $b$ cannot be both $1$ at the same position, the set $S$ is partitioned into three disjoint subsets (see figure \ref{figkkll})
\begin{equation}
S=S_0 \cup A_1 \cup B_1,
\end{equation}
where
\begin{eqnarray}
S_0 &=& \{i\in S \,|\, a_i=0,\; b_i=0\},  \nonumber\\
A_1 &=& \{i\in S \,|\, a_i=1,\; b_i=0\},  \nonumber\\
B_1 &=& \{i \in S \,|\, a_i=0,\; b_i=1\}.
\end{eqnarray}
Obviously, $|A_1|=|a|\leq n_A$ and $|B_1|=|b|\leq  n_{\bar{A}}$.

The number of terms $\#(k,k',l,l')$ is given by the number of bipartitions $(A,\bar{A})$ with $|A|=n_A$ such that
\begin{equation}
A_1\subset A \quad \mathrm{and}\quad B_1\subset \bar{A}.
\end{equation}
Since $A\cap\bar{A}=\emptyset$, parties $A$ and $\bar{A}$ contend for
$S_0$, namely
\beq
A=A_1\cup (S_0\cap A)
\quad \mbox{and}\quad \bar{A} = B_1\cup (S_0\cap \bar{A}).
\eeq
Thus, the number of bipartition is the number of ways of
picking $|A\backslash A_1|$ unordered outcomes from
$|S_0|$ possibilities. Since
$|A\backslash A_1|=|A|-|A_1|=n_A- |a|$ and
$|S_0|=|S|-|A_1|-|B_1|=n-|a|-|b|$, one gets
\barr
\label{eq:deltacounting}
\#(k,k',l,l')=\left(\begin{array}{c}
     n-|a|-|b|    \\
      {n_A}-|a|
\end{array} \right) .
\earr
Substituting Eq.\ (\ref{eq:deltacounting}) into
Eq.\ (\ref{eq:deltaappendix}) and by defining the binomial
coefficient to be identically zero when its arguments are negative,
we notice that the characteristic functions in (\ref{eq:deltaappendix})
yield always one, and obtain Eq.\ (\ref{eq:deltag}).

\section{}\label{sec:appendixb}

\begin{figure}
\begin{center}
\includegraphics[width=0.6\textwidth]{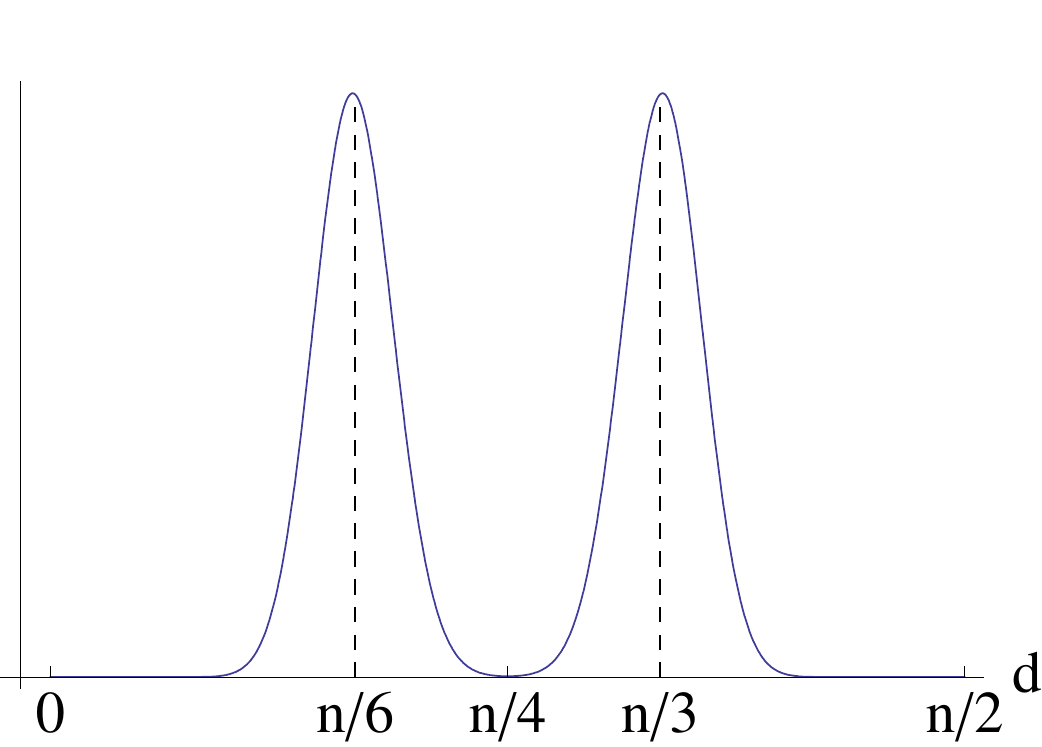}
\caption{(Color online) $d$-dependence of the
terms $\left(\!\!\begin{array}{c} n_A    \\ d \end{array}\!\!\right)
\left(\!\!\begin{array}{c} n_{\bar{A}}    \\ d \end{array}\!\!\right)
2^{n/2}\left[4^{n/4-d}+4^{-(n/4-d)}\right]$ in the sum (\ref{eq:f2app}). } \label{fig:saddle}
\end{center}
\end{figure}

We derive here the asymptotic (for large $N$) behavior of the
function $f_2(N)$ defined in Eq.\ (\ref{eq:f2}).
Let us define the \emph{distance} between bipartitions $(A,\bar{A})$
and $(B,\bar{B})$ as the number of qubits belonging to $A$ and not
to $B$
\beq
d=|A\cap\bar{B}|=|B\cap\bar{A}|\in[0,n_A].
\eeq
The number of pairs of bipartitions at a distance $d$ is
\barr
n_d=\left(\begin{array}{c}
     n    \\
    n_A
\end{array}\right)
\left(\begin{array}{c}
    n_A    \\
     d
\end{array}\right)
\left(\begin{array}{c}
  n_{\bar{A}}    \\
     d
\end{array}\right).
\label{eq:nd}
\earr
Therefore the sum over the bipartitions can be rewritten as a sum
over $d$
 \barr \sum_{|A|,|B|=n_A}[\cdots]= \left(\begin{array}{c}
     n    \\
    n_A
\end{array}\right)\sum_{d=0}^{n_A}
\left(\begin{array}{c}
    n_A    \\
     d
\end{array}\right)
\left(\begin{array}{c}
  n_{\bar{A}}    \\
     d
\end{array}\right)[\cdots].\earr
Let us consider for instance
\barr
\fl \quad \sum_{k,l\in {\mathbb{Z}}_2^n}{g(k,l)}^2
&=&
\frac{1}{N}\sum_{k\in {\mathbb{Z}}_2^{4n}}{\Delta(k_1,k_2;k_3,k_4)}^2\nonumber\\
&=&\frac{1}{N} \left(\!\!\begin{array}{c}
  n    \\
     n_{A}
\end{array}\!\!\right)^{\!\!\!-2}\sum_{k}\sum_{|A|,|B|=n_A} \frac{1}{4}\Big( \delta(k;A) \delta(k;B) + \delta(k;A) \delta(k;\bar{B})
\nonumber\\
& &
\qquad\qquad\qquad\qquad\qquad\quad +  \delta(k;\bar{A}) \delta(k; B) + \delta(k;\bar{A}) \delta(k;\bar{B})
\Big)
\nonumber\\
&=&\frac{1}{N} \left(\!\!\begin{array}{c}
  n    \\
     n_{A}
\end{array}\!\!\right)^{\!\!\!-2}\sum_{k}\sum_{|A|,|B|=n_A} \frac{1}{2}\Big( \delta(k;A) \delta(k;B) + \delta(k;A) \delta(k;\bar{B}) \Big)  ,
\earr
where
\begin{equation}
\delta(k,k',l,l';A)=  \delta_{k_{A}, l'_{A}} \delta_{k'_{A},l_{A}}
\delta_{k_{\bar{A}}, l_{\bar{A}}} \delta_{k'_{\bar{A}}, l'_{\bar{A}}}
\end{equation}
Let us start by showing that
\begin{equation}
\fl h(A,B)=\sum_{k} \delta(k;A) \delta(k; B)= \sum_{k,k',l,l'} \delta_{k_{A}, l'_{A}} \delta_{k'_{A},l_{A}} \delta_{k_{B}, l'_{B}} \delta_{k'_{B},l_{B}}
\delta_{k_{\bar{A}}, l_{\bar{A}}} \delta_{k'_{\bar{A}}, l'_{\bar{A}}}
\delta_{k_{\bar{B}}, l_{\bar{B}}} \delta_{k'_{\bar{B}}, l'_{\bar{B}}}
\end{equation}
depends only on $d=|A\cap \bar{B}|$.
When $A=B$, i.e.\ $d=0$,
\begin{equation}
\fl \qquad  h(A,A)=\sum_{k} \delta(k;A)^2= \sum_{k} \delta(k;A)= \sum_{k,k',l,l'} \delta_{k_{A}, l'_{A}} \delta_{k'_{A},l_{A}}\delta_{k_{\bar{A}}, l_{\bar{A}}} \delta_{k'_{\bar{A}}, l'_{\bar{A}}} =N^2,
\end{equation}
while, when $d\neq 0$, we get
\begin{equation}
 h(A,B)= \left(\frac{N}{2^d}\right)^2 = 4^{-d} N^2.
\end{equation}
Therefore, we get
\begin{eqnarray}
\fl & & 
\sum_{k}  \sum_{|A|,|B|=n_A} \delta(k;A) \delta(k;B)
= \sum_{|A|,|B|=n_A} h(A,B)\nonumber\\
& & =  \sum_{0\leq d\leq n_A} n_d\;  4^{-d} N^2 =
\left(\!\!\begin{array}{c} n    \\ n_{A} \end{array}\!\!\right)
\!\sum_{0\leq d\leq n_A}\!\! 4^{-d}
\left(\!\!\begin{array}{c} n_A    \\ d \end{array}\!\!\right)
\left(\!\!\begin{array}{c} n_{\bar{A}}    \\ d \end{array}\!\!\right).
\label{eq:app_bg2}
\end{eqnarray}
Analogously we find
\barr
\label{eq:app_bgg}
\sum_{k}  \sum_{|A|,|B|=n_A} \delta(k;A) \delta(k;\bar{B})
=\left(\!\!\begin{array}{c} n    \\ n_{A} \end{array}\!\!\right)
\!\sum_{0\leq d\leq n_A}\!\! 4^{d}
\left(\!\!\begin{array}{c} n_A    \\ d \end{array}\!\!\right)
\left(\!\!\begin{array}{c} n_{\bar{A}}    \\ d \end{array}\!\!\right).
\earr
Putting together Eqs.\ (\ref{eq:app_bg2})-(\ref{eq:app_bgg})
we get
\barr
 f_2(N)&=& 4 \sum_{k,l}g(k,l)^2\nonumber\\
&=& 2
\left(\!\!\begin{array}{c} n    \\ n_{A} \end{array}\!\!\right)^{\!\!\!-1}
\!\!\!\sum_{0\leq d\leq n_A}\!\!
\left(\!\!\begin{array}{c} n_A    \\ d \end{array}\!\!\right)
\left(\!\!\begin{array}{c} n_{\bar{A}}    \\ d \end{array}\!\!\right)
(4^{n/2-d}+4^d)=
\nonumber\\
&=& 2
\left(\!\!\begin{array}{c} n    \\ n_{A} \end{array}\!\!\right)^{\!\!\!-1}
\!\!\!\sum_{0\leq d\leq n_A}\!\!
\left(\!\!\begin{array}{c} n_A    \\ d \end{array}\!\!\right)
\left(\!\!\begin{array}{c} n_{\bar{A}}    \\ d \end{array}\!\!\right)
2^{n/2}\left[4^{n/4-d}+4^{-(n/4-d)}\right].
\label{eq:f2app}
\earr
We notice that the terms in the summation strongly depend on
the ratio $2d/n$. In the limit $n\rightarrow+\infty$ only the
terms with $d=n/6$ and $d=n/3$ give a significant contribution to
the summation (see Fig.\ \ref{fig:saddle}).
Let us consider the case of even $n$ (in the thermodynamic limit the result for an odd number of qubits is the same)
\beq
\label{eq:balancedbip}
n_A=n_{\bar{A}}=n/2.
\eeq
By Stirling's approximation $n!\sim (n/e)^n \sqrt{2\pi n}$ (for $n$ large) and by defining the new variable
\beq
x=\frac{2d}{n},
\eeq
after a straightforward calculation we obtain
\barr
 f_2(N) &\sim& 2 \sqrt{\frac{\pi n}{2}}2^{-n}\sum_{d}\frac{1}{\pi n x \left(1-x\right)}\exp{\left\{n S(x)\right\}}2^{n/2}\left[4^{\frac{n}{2}(\frac{1}{2}-x)}+4^{-\frac{n}{2}(\frac{1}{2}-x)}\right]\nonumber\\
&\sim&\sqrt{2\pi n} \int_0^1\d x \frac{1}{2\pi x(1-x)}\nonumber\\
&\times&\left[\exp{\left\{n\left[S(x)-x\ln 2\right]\right\}}+2^{-n}\exp{\left\{n\left[S(x)+x\ln 2\right]\right\}}\right],
\earr
where
\beq
S(x)=-x\ln x-(1-x)\ln(1-x)
\eeq
is the Shannon entropy.
Using the saddle point approximation in the integrand we get
\barr
\label{eq:saddlepoint}
f_2(N) &\sim& \sqrt{2\pi n} \int_0^1\d x \frac{9}{4\pi }\nonumber\\&\times&
\left[\exp{\left\{n\left[S\left(\frac{1}{3}\right)-\frac{1}{3}\ln 2+\frac{1}{2}S''\left(\frac{1}{3}\right){(x-\frac{1}{3})}^2\right]\right\}}+\right.\nonumber\\
&+&\left.2^{-n}\exp{\left\{n\left[S\left(\frac{2}{3}\right)+\frac{2}{3}\ln 2+\frac{1}{2}S''\left(\frac{2}{3}\right){(x-\frac{2}{3})}^2\right]\right\}}\right]\nonumber\\
&=& \sqrt{2\pi n} \int_0^1\d x \frac{9}{4\pi }\exp\left\{n\ln\frac{3}{2}\right\}\nonumber\\&\times&\left[\exp\left\{-n \frac{9}{4} {\left(x-\frac{1}{3}\right)}^2\right\}+\exp\left\{-n \frac{9}{4} {\left(x-\frac{2}{3}\right)}^2\right\}\right]\nonumber\\
&\sim& \frac{9}{\sqrt{2\pi }} {\left(\frac{3}{2}\right)}^n \int_{-\infty}^{+\infty} \d x \exp\left\{-\frac{9}{4}x^2 \right\}=3 \sqrt{2} N^\alpha,
\earr
where
\beq
\alpha=\log_2 3-1 \simeq 0.584963.
\eeq
This is the asymptotic expression
(\ref{eq:f2}) used in Eq.\ (\ref{eq:secondocum}).

\section{}
\label{sec:appendixe}

We evaluate here the asymptotic behavior of the function $f_3^{(0)}(N)$ defined in (\ref{eq:f30}).
By using the definition (\ref{eq:gdef}), (\ref{eq:f30}) can be written
\barr\label{eq:appendixf30}
f_3^{(0)}(N)&=&\sum_{k_1,k_2,k_3}g(k_1,k_2)\;g(k_3,k_2)\;g(k_1\oplus k_3,k_2)\nonumber\\
&=&\sum_{k_1,k_2,k_3}\delta_{k_1\wedge k_2,0}\delta_{k_2\wedge k_3,0}\delta_{k_2\wedge (k_1\oplus k_3),0}\nonumber\\
&\times&\hat{g}(|k_1|,|k_2|)\;\hat{g}(|k_3|,|k_2|)\;\hat{g}(|k_1\oplus k_3|,|k_2|)\nonumber\\
&=&\sum_{s_0,s_1,s_2,s_3}f(s_0,s_1,s_2,s_3)\hat{g}(s_1,s_2)\;\hat{g}(s_3,s_2)\;\hat{g}(s_1+s_3-2s_0,s_2),\nonumber\\
\earr
where
\barr
f(s_0,s_1,s_2,s_3)=\sum_{k_1,k_2,k_3}\delta_{k_1\wedge k_2,0}\delta_{k_2\wedge k_3,0}\delta_{s_1,|k_1|}\delta_{s_2,|k_2|}\delta_{s_3,|k_3|}\delta_{s_0,|k_1\wedge k_3|}
\label{eq:fs0s1s2s3}
\earr
and we have used 
\barr
k_1\wedge k_2=0,\quad k_2\wedge k_3=0 \quad \Rightarrow \quad k_2\wedge (k_1\oplus k_3)=0,\\
|k_1\oplus k_3|=|k_1|+|k_3|-2|k_1\wedge k_3|.
\earr
It is straightforward to count the number of terms in (\ref{eq:fs0s1s2s3}) and obtain
\barr
f(s_0,s_1,s_2,s_3)&=&
\left(\!\!\begin{array}{c}
         n    \\
      s_2
\end{array}\!\!\right)
\left(\!\!\begin{array}{c}
         n-s_2    \\
      s_1
\end{array} \!\!\right)
\left(\!\!\begin{array}{c}
         s_1    \\
      s_0
\end{array} \!\!\right)
\left(\!\!\begin{array}{c}
         n-s_2-s_1    \\
      s_3-s_0
\end{array} \!\!\right)\nonumber\\
&=&\frac{n!}{s_2!s_0!(s_1-s_0)!(s_3-s_0)!(n-s_2-s_1-s_3+s_0)!}.
\earr
By substituting $s_1\to s_1+s_0$ and $s_3 \to s_3+s_0$ in Eq.\ (\ref{eq:appendixf30}) we obtain
\barr
f_3^{(0)}(N)&=&\sum_{s_0,s_1,s_2,s_3}\frac{n!}{s_0!s_1!s_2!s_3!(n-s_1-s_2-s_3-s_0)!}\nonumber\\
&\times&\hat{g}(s_1+s_0,s_2)\;\hat{g}(s_3+s_0,s_2)\;\hat{g}(s_1+s_3,s_2)\nonumber\\
&=&\sum_s\left(\!\!\begin{array}{c}
         n   \\
      s_0,s_1,s_2,s_3
\end{array} \!\!\right)\hat{g}(s_1+s_0,s_2)\;\hat{g}(s_3+s_0,s_2)\;\hat{g}(s_1+s_3,s_2),\nonumber\\
\earr
where 
\barr
\left(\!\!\begin{array}{c}
         n   \\
      i_1,i_2,...,i_k
\end{array} \!\!\right)=\frac{n!}{\Pi_{j=1}^k (i_j!)(n-\sum_{j=1}^k i_j)!}
\label{eq:multinomdef}
\earr
denotes the multinomial coefficient. A relabeling of the dummy variables yields
\begin{equation}
\fl \qquad f_3^{(0)}(N) = \sum_s\left(\!\!\begin{array}{c}
         n   \\
      s_0,s_1,s_2,s_3
\end{array} \!\!\right)\hat{g}(s_1+s_2,s_0)\;\hat{g}(s_2+s_3,s_0)\;\hat{g}(s_3+s_1,s_0).
\end{equation}

From the definition (\ref{eq:hatg}) one easily get
\begin{equation}
\hat{g}(s,t)=
\frac{1}{2} \left(\!\! \begin{array}{c}
         n   \\
      s,t
\end{array} \!\!\right)^{\!\!-1}
\left[\left(\!\!\begin{array}{c}
         n_A   \\
      s
\end{array} \!\!\right)
\left(\!\!\begin{array}{c}
         n_{\bar{A}}   \\
        t
\end{array} \!\!\right) +
\left(\!\!\begin{array}{c}
         n_A   \\
      t
\end{array} \!\!\right)
\left(\!\!\begin{array}{c}
         n_{\bar{A}}   \\
        s
\end{array} \!\!\right)
 \right] .
\label{eq:hatgmult} 
\end{equation}
Therefore, for $n_A=n_{\bar{A}}=n/2$  we finally obtain
\begin{equation}
\fl \quad f_3^{(0)}(N)=\sum_s\left(\!\!\begin{array}{c}
         n   \\
      s_0,s_1,s_2,s_3
\end{array} \!\!\right)
\left(\!\!\begin{array}{c}
         n/2   \\
      s_0
\end{array} \!\!\right)^{\!\!3}
\prod_{1\leq i\leq 3}
\left(\!\!\begin{array}{c}
         n   \\
     s_0, s_i+s_{i+1}
\end{array} \!\!\right)^{\!\!-1}
\left(\!\!\begin{array}{c}
         n/2   \\
      s_i+s_{i+1}
\end{array}\!\! \right).
\end{equation}

Now, by  using the Stirling approximation and scaling the variables
\beq\label{eq:rescalvar}
\sigma_0=\frac{s_0}{n}, \;\sigma_1=\frac{s_1}{n},\;\sigma_2=\frac{s_2}{n},\; \sigma_3=\frac{s_3}{n}
\eeq
we obtain, after some algebra, the asymptotic form
\begin{equation}
\label{eq:asymf30}
f_3^{(0)}(N)  \sim \frac{1}{(2\pi n)^2}\sum_s A (\sigma_0,\sigma_1,\sigma_2,\sigma_3)\, \exp\{nS(\sigma_0,\sigma_1,\sigma_2,\sigma_3)\}.
\end{equation}
In Eq.\ (\ref{eq:asymf30}) we have set (with the implicit convention that the indices are cyclical)
\barr
\label{eq:coefficientA}
\fl \quad A = \sqrt{\frac{\prod_{i=1}^3 (1-\sigma_0-\sigma_i-\sigma_{i+1})}{\sigma_0(1-2\sigma_0)^3(1-\sigma_0-\sigma_1-\sigma_2-\sigma_3)
\prod_{i=1}^3 \sigma_i (1-2\sigma_i-2\sigma_{i+1})}}
\earr
and
\barr
S(\sigma_0,\sigma_1,\sigma_2,\sigma_3)&=&S_4(\sigma_0,\sigma_1,\sigma_2,\sigma_3)-S_2(\sigma_0,\sigma_1+\sigma_2)-S_2 (\sigma_0,\sigma_2+\sigma_3)\nonumber\\
&-&S_2 (\sigma_0,\sigma_3+\sigma_1) +\frac{3}{2}S_1(2\sigma_0)+\frac{1}{2}S_1(2\sigma_1+2\sigma_2)\nonumber\\
&+&\frac{1}{2}S_1(2\sigma_2+2\sigma_3)+\frac{1}{2}S_1(2\sigma_3+2\sigma_1),
\earr
with
\barr\label{eq:appendixentropy}
S_n(x_1,\dots,x_n)= - \sum_{i=1}^n x_i\log x_i - \Big(1-\sum_{i=1}^n x_i\Big) \log\!\Big(1-\sum_{i=1}^n x_i\Big).
\earr
By noting that
\begin{equation}
S_n(x_1,\dots,x_n) = \sum_{i=1}^n x_i \frac{\partial S_n}{\partial x_i} - \log\!\Big(1-\sum_{i=1}^n x_i\Big),
\label{eq:quasihom}
\end{equation}
one easily gets 
\begin{equation}
 S= \sigma_0 \frac{\partial S}{\partial \sigma_0} + \sum_{i=1}^3 \sigma_i \frac{\partial S}{\partial \sigma_i}  + S_0 ,
\label{eq:quasihom1}
\end{equation}
with
\begin{equation}
\fl \qquad S_0
= \frac{1}{2} \log \frac{\prod_{i=1}^3 (1-\sigma_0-\sigma_i-\sigma_{i+1})^2}{(1-\sigma_0-\sigma_1-\sigma_2-\sigma_3)^2 (1-2\sigma_0)^3 \prod_{i=1}^3 (1-2\sigma_i-2\sigma_{i+1}) }.
\end{equation}

In the limit $n\to\infty$ the main contribution comes from the saddle point $(\sigma_0^*,\sigma_1^*,\sigma_2^*,\sigma_3^*)$, solution to the system
\barr
\frac{\partial S}{\partial \sigma_i}=0, \mbox{with} \;\; i=0,1,2,3,
\earr
that reads
\barr
(1-\sigma_0-\sigma_1-\sigma_2-\sigma_3)\left(1-2\sigma_0\right)^3= 8 \sigma_0 \prod_{i=1}^3 (1-\sigma_0-\sigma_i-\sigma_{i+1}),
\earr
\barr
\fl & &(1-\sigma_0-\sigma_1-\sigma_2-\sigma_3) \left(1-2\sigma_i-2\sigma_{i+1}\right)\left(1-2\sigma_i-2\sigma_{i+2}\right)\nonumber\\
& & \quad = 4 \sigma_i
(1-\sigma_0-\sigma_i-\sigma_{i+1})(1-\sigma_0-\sigma_i-\sigma_{i+2}),
\label{eq:saddlepointeqs}
\earr
with $i=1,2,3$.
In the limit $n\to\infty$ we get
\barr
\fl \quad f_3^{(0)}(N) &\sim& \left(\frac{n}{2\pi}\right)^2 A^* \e^{n S_0^*} \int_{\mathbb{R}^4} \exp\left(\frac{n}{2}\sum_{i,j=0}^3\frac{\partial^2S^*}{\partial\sigma_i\partial\sigma_j}(\sigma_i-\sigma_i^*)(\sigma_j-\sigma_j^*)\right) \d\sigma_0 \d\sigma_1 \d\sigma_2 \d\sigma_3 \nonumber\\
\fl \quad &=& A^* \det\left(\frac{\partial^2S^*}{\partial\sigma_i\partial\sigma_j}\right)^{-1/2}   \exp\left(n S_0^*\right)
\label{eq:asymf30sym}
\earr
where the starred functions $A^*$, $S^*$, and $\partial^2 S^* /\partial\sigma_i\partial\sigma_j$ are evaluated at the saddle point $(\sigma_0^*,\sigma_1^*,\sigma_2^*,\sigma_3^*)$.

  The symmetry of the equations suggests to look at a symmetric solution of (\ref {eq:saddlepointeqs}) with
\beq
\sigma_i=\sigma \quad \mbox{with} \;\; i=1,2,3,
\eeq
which yields
\barr
(1-\sigma_0-3\sigma)(1-2\sigma_0)^3=8 \sigma_0(1-\sigma_0-2\sigma)^3, \nonumber \\
(1-\sigma_0-3\sigma)(1-4\sigma)^2=4 \sigma(1-\sigma_0-2\sigma)^2.
\earr
We get
\barr
\fl\quad  A^* = \sqrt{\frac{ (1-\sigma_0^*-2\sigma^*)^3 }{\sigma_0^* \sigma^{*3}(1-2\sigma_0^*)^3  (1-4\sigma^*)^3(1-\sigma_0^*-3\sigma^*)}}
 = \sqrt{\frac{1}{8\sigma_0^{*2} \sigma^{*3} (1-4\sigma^*)^3}},
\earr
\begin{equation}
\fl\quad  S_0^*= \frac{1}{2} \log \frac{ (1-\sigma_0^*-2\sigma^*)^6}{(1-\sigma_0^*-3\sigma^*)^2 (1-2\sigma_0^*)^3  (1-4\sigma^*)^3 }
= \frac{1}{2} \log \frac{ (1- 2\sigma_0^*)^3}{64\sigma_0^{*2} (1-4\sigma^*)^3 },
\end{equation}
and
\begin{eqnarray}
\fl\quad  \det\left(\frac{\partial^2S^*}{\partial\sigma_i\partial\sigma_j}\right) &=& 
\frac{ 
(1  - \sigma_0^*  -5 \sigma^* + 2\sigma^*(4 \sigma^{*}+  \sigma_0^*))^2}{\sigma_0^* \sigma^{*3}(1-2\sigma_0^*)  (1-4\sigma^*)^3(1-\sigma_0^*-3\sigma^*)(1-\sigma_0^*-2\sigma^*)^3}
\nonumber\\
& &\times
(1 - 2\sigma^* - \sigma_0^{*2} - 4 \sigma^{*2}  + 
   4 \sigma^* \sigma_0^*  (8 \sigma^* + 2 \sigma_0^{*}- 3))
\end{eqnarray}

The solution of the system that gives the largest contribution is
\barr\label{eq:solutionf31}
\sigma_0^*&=&\frac{13}{36}-\frac{13}{36 \sqrt[3]{197-18 \sqrt{113}}}-\frac{1}{36} \sqrt[3]{197-18 \sqrt{113}}\simeq 0.108955,
\nonumber\\
\sigma^*&=&-\frac{5043923}{144 \left(197-18 \sqrt{113}\right)^{7/3}}+\frac{158161 \sqrt{113}}{48 \left(197-18
   \sqrt{113}\right)^{7/3}}\nonumber\\
   && +\frac{980473}{72 \left(197-18 \sqrt{113}\right)^2}-\frac{2561 \sqrt{113}}{2 \left(197-18
   \sqrt{113}\right)^2}\nonumber\\
   &&-\frac{18119}{144 \left(197-18 \sqrt{113}\right)^{5/3}}+\frac{563 \sqrt{113}}{48 \left(197-18
   \sqrt{113}\right)^{5/3}} \nonumber \\ &\simeq &0.104767.
    \earr
Plugging these results into Eq.\ (\ref {eq:asymf30sym})  we get
\barr
f_3^{(0)}(N)
\sim c \; N^{5-\gamma}
\label{eq:asymf30bis}
\earr
where 
\begin{eqnarray}
\fl c &=& A^* \det\left(\frac{\partial^2S^*}{\partial\sigma_i\partial\sigma_j}\right)^{-1/2} 
\nonumber\\
\fl &=& \frac{ (1-\sigma_0^*-2\sigma^*)^3 }{1-2\sigma_0^*}
\nonumber\\
\fl & &\times \frac{1}{(1  - \sigma_0^*  -5 \sigma^* + 2\sigma^*(4 \sigma^{*}+  \sigma_0^*))\sqrt{1 - 2\sigma^* - \sigma_0^{*2} - 4 \sigma^{*2}  + 
   4 \sigma^* \sigma_0^*  (8 \sigma^* + 2 \sigma_0^{*}- 3)}}
\nonumber\\
\fl &\simeq&  1.05385,
\label{eq:cdef1}
\end{eqnarray} 
and
\beq
\label{eq:gammadef1}
\gamma = 5- S_0^* \log_2 \e=
5-\frac{1}{2}\log_2 \left[ \frac{(1-2\sigma_0^*)^3}{64 \sigma_0^{*2}(1-4\sigma^{*})^3}\right] 
\simeq 4.1583.
\eeq

\section{}
\label{sec:appendixe2}

We evaluate here the asymptotic behavior of the function $f_3^{(1)}(N)$
defined in (\ref{eq:f31}).
By using the definition (\ref{eq:gdef}) we have
\barr\label{eq:appendixf31}
 f_3^{(1)}(N)&=&\sum_{k_1,k_2,k_3}g(k_1,k_2\oplus k_3)\;g(k_2,k_1\oplus k_3)\;g(k_3,k_1\oplus k_2)\nonumber\\
 &=&\sum_{k_1,k_2,k_3}\delta_{k_1\wedge (k_2\oplus k_3),0}\delta_{k_2\wedge (k_1\oplus k_3),0}\delta_{k_3\wedge (k_1\oplus k_2),0}\nonumber\\
 & &\times \hat{g}(|k_1|,|k_2\oplus k_3|)\;\hat{g}(|k_2|,|k_1\oplus k_3|)\;\hat{g}(|k_3|,|k_1\oplus k_3|)\nonumber\\
&=&\sum_{s_0,s_1,s_2,s_3}h(s_0,s_1,s_2,s_3)\hat{g}(s_1,s_2+s_3-2s_0)\;\hat{g}(s_2,s_1+s_2-2s_0)
\nonumber\\
 & &\times \hat{g}(s_3,s_1+s_2-2s_0),
\earr
where
\barr
h(s_0,s_1,s_2,s_3)=\sum_{k_1,k_2,k_3}\delta_{s_1,|k_1|}\delta_{s_2,|k_2|}\delta_{s_3,|k_3|}\delta_{s_0,|k_1\wedge k_2|}\delta_{s_0,|k_1\wedge k_3|}\delta_{s_0,|k_2\wedge k_3|}\nonumber\\
\earr
and we have used 
\barr
&&k_1\wedge (k_2\oplus k_3)=0,\quad k_2\wedge (k_1\oplus k_3)=0,\quad k_3\wedge (k_1\oplus k_2)=0
\nonumber\\ 
&& \Rightarrow k_1\wedge k_2=k_1\wedge k_3=k_2\wedge k_3,
\nonumber \\
&&|k_i\oplus k_j|=|k_i|+|k_j|-2|k_i\wedge k_j|\quad\forall\;i,j=1,2,3.
\earr
We find
\barr
\fl \quad h(s_0,s_1,s_2,s_3)&=&\frac{n!}{s_0!(s_1-s_0)!(s_2-s_0)!(s_3-s_0)!(n-s_2-s_1-s_3+2s_0)!}.
\earr
Using the substitution $s_1\to s_1+s_0$, $s_2\to s_2+s_0$ and $s_3\to s_3+s_0$ in Eq.\ (\ref{eq:appendixf31}) we obtain
\barr
\fl \quad f_3^{(1)}(N)
= \sum_s
\left(\!\!\begin{array}{c}
         n   \\
      s_0,s_1,s_2,s_3
\end{array} \!\!\right)
\hat{g}(s_1+s_0,s_2+s_3)\;\hat{g}(s_2+s_0,s_1+s_3)\;\hat{g}(s_3+s_0,s_1+s_2),
\nonumber\\
\earr
in terms of the multinomial coefficient (\ref{eq:multinomdef}).
Using (\ref{eq:hatgmult} ) for $n_A=n_{\bar{A}}=n/2$  we finally obtain
\begin{eqnarray}
\fl \quad f_3^{(1)}(N)=\sum_s\left(\!\!\begin{array}{c}
         n   \\
      s_0,s_1,s_2,s_3
\end{array} \!\!\right)
\prod_{1\leq i\leq 3}
\left(\!\!\begin{array}{c}
         n   \\
     s_0+s_i, s_{i+1}+s_{i+2}
\end{array} \!\!\right)^{\!\!-1}
\left(\!\!\begin{array}{c}
         n/2   \\
      s_0+s_i
\end{array} \!\!\right)
\left(\!\!\begin{array}{c}
         n/2   \\
      s_i+s_{i+1}
\end{array}\!\! \right).
\nonumber\\
\end{eqnarray}

Using the Stirling approximation and Eq.\ (\ref{eq:rescalvar}) we get
\begin{equation}
\label{eq:asymf31}
f_3^{(1)}(N)  \sim \frac{1}{(2\pi n)^2}\sum_s A (\sigma_0,\sigma_1,\sigma_2,\sigma_3)\, \exp\{nS(\sigma_0,\sigma_1,\sigma_2,\sigma_3)\},
\end{equation}
where
\barr
\fl \qquad A(\sigma_0,\sigma_1,\sigma_2,\sigma_3)=\frac{(1-\sigma_0-\sigma_1-\sigma_2-\sigma_3)}{\sqrt{\sigma_0\sigma_1\sigma_2\sigma_3
\prod_{1\leq i\leq 3}
(1-2\sigma_0-2\sigma_i)(1-2\sigma_{i+1}-2\sigma_{i+2})}}
\earr
and
\barr
\fl \qquad S(\sigma_0,\sigma_1,\sigma_2,\sigma_3)&=&S_4(\sigma_0,\sigma_1,\sigma_2,\sigma_3)-S_2(\sigma_0+\sigma_1,\sigma_2+\sigma_3)\nonumber\\
\fl & & - S_2(\sigma_0+\sigma_2,\sigma_3+\sigma_1)
-S_2(\sigma_0+\sigma_3,\sigma_1+\sigma_2)\nonumber\\
\fl & &+\frac{1}{2}S_1(2\sigma_0+2\sigma_1)+\frac{1}{2}S_1(2\sigma_0+2\sigma_2) +\frac{1}{2}S_1(2\sigma_0+2\sigma_3)\nonumber\\
\fl & &+\frac{1}{2}S_1(2\sigma_1+2\sigma_2)+\frac{1}{2}S_1(2\sigma_2+2\sigma_3)+\frac{1}{2}S_1(2\sigma_3+2\sigma_1),
\earr
with the entropies defined in Eq.\ (\ref{eq:appendixentropy}). By (\ref{eq:quasihom}) one gets (\ref{eq:quasihom1}) with
\begin{equation}
S_0 =\frac{1}{2} \log \frac{(1-\sigma_0-\sigma_1-\sigma_2-\sigma_3)^4}{
\prod_{1\leq i\leq 3}
(1-2\sigma_0-2\sigma_i)(1-2\sigma_{i+1}-2\sigma_{i+2})}.
\end{equation}

In the limit $n\rightarrow+\infty$ we can use the saddle point approximation. The saddle point is solution to the set of equations
\barr
\frac{\partial S}{\partial \sigma_i}=0, \quad \mbox{with} \;\; i=0,1,2,3,
\earr
that reads
\begin{eqnarray}
\fl \quad (1-2\sigma_i-2\sigma_{i+1})(1-2\sigma_i-2\sigma_{i+2})(1-2\sigma_i-2\sigma_{i+3})=8\sigma_i (1-\sigma_0-\sigma_1-\sigma_2-\sigma_3)^2.
\nonumber\\
\end{eqnarray}
The symmetric solution, $\sigma_i=\sigma$ for all $i$, corresponds
to the largest contribution and is given by
\beq
\sigma*=\frac{1}{12}.
\eeq
As in (\ref{eq:asymf30sym}), in the limit $n\to\infty$ we get
\begin{equation}
f_3^{(1)}(N) \sim A^* \det\left(\frac{\partial^2S^*}{\partial\sigma_i\partial\sigma_j}\right)^{-1/2}   \exp\left(n S_0^*\right),
\end{equation}
with
\begin{equation}
\fl\qquad A^*=\frac{1}{\sigma^{*2}
(1-4\sigma^*)^2} =324,
\qquad
 \det\left(\frac{\partial^2S^*}{\partial\sigma_i\partial\sigma_j}\right)= \frac{1}{\sigma^{*4}
(1-4\sigma^*)^4} =324^2,
\end{equation}
and
\begin{equation}
S_0^*= \log \frac{1}{(1-4\sigma^*)}= \log \frac{3}{2}.
\end{equation}
We finally get
\barr
f_3^{(1)}(N)\sim
\e^{n \log (3/2)}= N^\alpha, \qquad \mbox{for} \quad N\to\infty,
\label{eq:asymf31bis}
\earr
with
\beq
\alpha={\log_2 3-1}  \simeq 0.584963.
\eeq



\end{document}